\documentclass[12pt]{article}
\usepackage{a4}
\usepackage{epsfig}
\def\Journal#1#2#3#4{{#1} {\bf #2}, #3 (#4)}

\def\PLB{{\em Phys. Lett.}  B}
\def\PRL{\em Phys. Rev. Lett.}
\def\PRD{{\em Phys. Rev.} D}
\def\PRP{{\em Phys. Reports}}
\def\ZPC{{\em Z. Phys.} C}
\def\COMP{{\em Comput. Phys. Commun.} }

\begin{document}
\newcommand{\newc}{\newcommand}
\newc{\R}{$R$}
\newc{\charginom}{M_{\tilde \chi}^{+}}
\newc{\mue}{\mu_{\tilde{e}_{iL}}}
\newc{\mud}{\mu_{\tilde{d}_{jL}}}
\newc{\barr}{\begin{eqnarray}}
\newc{\earr}{\end{eqnarray}}
\newc{\beq}{\begin{equation}}
\newc{\eeq}{\end{equation}}
\newc{\ra}{\rightarrow}
\newc{\lam}{\lambda}
\newc{\eps}{\epsilon}
\newc{\gev}{\,GeV}
\newc{\tev}{\,TeV}
\newc{\eq}[1]{(\ref{eq:#1})}
\newc{\eqs}[2]{(\ref{eq:#1},\ref{eq:#2})}
\newc{\etal}{{\it et al.}\ }
\newc{\eg}{{\it e.g.}\ }
\newc{\ie}{{\it i.e.}\ }
\newc{\nonum}{\nonumber}
\newc{\lab}[1]{\label{eq:#1}}
\newc{\dpr}[2]{({#1}\cdot{#2})}
\newc{\gsim}{\stackrel{>}{\sim}}
\newc{\lsim}{\stackrel{<}{\sim}}
\begin{titlepage}
\begin{flushright}
{ETHZ-IPP  PR-98-09} \\
{December 4, 1998}\\
\end{flushright}
\begin{center}
{\bf \LARGE Experimental Challenges at the LHC} \\
\end{center}

\smallskip \smallskip \bigskip
\begin{center}
{\Large  Felicitas Pauss and Michael Dittmar }
\end{center}
\bigskip
\begin{center}
Institute for Particle Physics (IPP), ETH Z\"{u}rich, \\
CH-8093 Z\"{u}rich, Switzerland
\end{center}

\bigskip
\begin{abstract}
\noindent 
The Large Hadron Collider (LHC) at CERN will provide proton-proton
collisions at a centre-of-mass energy of 14 TeV with a design luminosity
of 10$^{34}$cm$^{-2}$s$^{-1}$. 
The exploitation of the rich
physics potential offered by the LHC will be illustrated using the expected
performance of the two general--purpose detectors ATLAS and CMS. \\ 
The detector design requirements necessary to extract the physics
under the challenging experimental conditions  at the LHC are discussed.
This is followed by an 
analysis of search methods for the Higgs sector and the
detection of supersymmetric particles.
\end{abstract}
\vspace{2cm}

\begin{center}
{\it Lectures given at } \\
ASI Summer School, St.Croix, 1998
\end{center}
\end{titlepage}

\section{Introduction}
The Standard Model (SM) of particle physics -- the theory of electroweak and
strong forces -- provides a remarkably successful theoretical 
picture~\cite{maria}. 
The SM has been tested rigorously at LEP, the Tevatron and the 
linear collider at SLAC~\cite{nodulman}. 
The four LEP experiments
have already given a definitive answer to the number of fundamental
building blocks of matter: there exist three families of
quarks and leptons with a light neutrino. 

One of the key questions in particle physics today is the
origin of the spontaneous symmetry breaking mechanism. 
The electroweak sector of the SM postulates that the Higgs 
mechanism is responsible for this symmetry breaking, 
and predicts a scalar Higgs boson.
Introducing this Higgs boson in the SM 
allows the masses of all particles to be expressed in terms 
of their couplings to the Higgs.
In order to complete the SM prediction we therefore have to establish 
experimentally the existence of the {\em last} missing element: 
the {\em Higgs} boson~\cite{Higgshunter}.

Even though the SM describes existing data very well, and even 
if it successfully passes further tests, we know that this Model 
is incomplete, as it supplies no answer to some fundamental questions. 
One problem of the SM is the instability of the mass of an elementary 
scalar, such as the Higgs boson, under radiative corrections in 
the presence of a high scale, like for example the Planck
scale ($\approx 10^{19}$ GeV). These divergences disappear in
{\em Supersymmetry} (SUSY)~\cite{nilles}, because of cancellations between the virtual
effects of SM particles and their supersymmetric partners,
which are introduced 
to every known fermion and boson of the SM. 
Furthermore, SUSY must be a 
broken symmetry because known particles have no super partner 
of the same mass. These must be heavier, and are therefore not yet discovered.

Another problem originates 
from  extrapolating the coupling strength of the fundamental forces measured
at mass scales of a few 100 GeV to energy scales
relevant for cosmology, i.e. 
energies of about $10^{15}$ to $10^{19}$ GeV.  
Performing this extrapolation within the SM does not lead to 
unification of forces at very high scales. Introducing however
SUSY unification of the electromagnetic, weak 
and strong forces at the GUT scale ($\approx 10^{15}$ GeV) is  
predicted which is consistent with  a SUSY mass scale of $\mathcal{O}$(TeV).

It is possible that the Higgs boson is an elementary particle 
as predicted in the SM and its supersymmetric extension. 
Alternatives to a fundamental scalar Higgs involve {\em new 
strong forces}. In models without a scalar Higgs, 
the W and Z masses could then be 
due to a dynamical symmetry breaking~\cite{alternative}.
In such a scenario the symmetry breaking could lead to a 
strong interaction between the longitudinal components of the 
intermediate vector bosons (W$_{L}$, Z$_{L}$). This strong interaction 
may be resonant or not. Resonances may occur in analogy 
with $\pi\pi$--scattering, which leads to spin=1 $\rho$--like states, 
or spin=0 very broad resonances, expected to be in the TeV mass range.

In spite of the impressive success of the SM there is a general consensus
that the SM is not the ultimate description of nature, and that new phenomena 
should manifest themselves in the energy region of order 1 TeV. 
The Large Hadron Collider (LHC), operating at a centre--of--mass
energy of 14 TeV with a design luminosity of 10$^{34}$cm$^{-2}$s$^{-1}$,  
will be the first machine 
to probe parton--parton collisions 
directly at energies $\approx$ 1 TeV~\cite{lhcref}. Such energies will be 
essential to address, for example, the questions of the origin of 
spontaneous symmetry breaking.

Currently no experimental evidence exists for any exotic new phenomena,
therefore  our discussion on search strategies at the LHC 
should  be quite general. We follow however today's
 fashion and focus 
mainly on  the  detector requirements 
necessary for future successful searches for the Higgs
and for Supersymmetry.
The presented ideas and methods should nevertheless provide 
a good guidance for more ``exotic''  searches. 
%
%
\section {Present Experimental Status of the Standard Model and Beyond}
In the following we summarise briefly the present status of
physics topics relevant for LHC and speculate about what one might
know from future experimental results before the start--up of the LHC,
presently foreseen in the year 2005.

\subsection {The Higgs sector}
It is well known that the value of the Higgs mass is not predictable
within the SM. On the other hand, the Higgs cannot be too heavy, otherwise
the perturbative regime breaks down, and this leads to an upper bound 
on the Higgs mass of about 1000 GeV.
\begin{figure}[htb]
\begin{center}
\epsfig{file=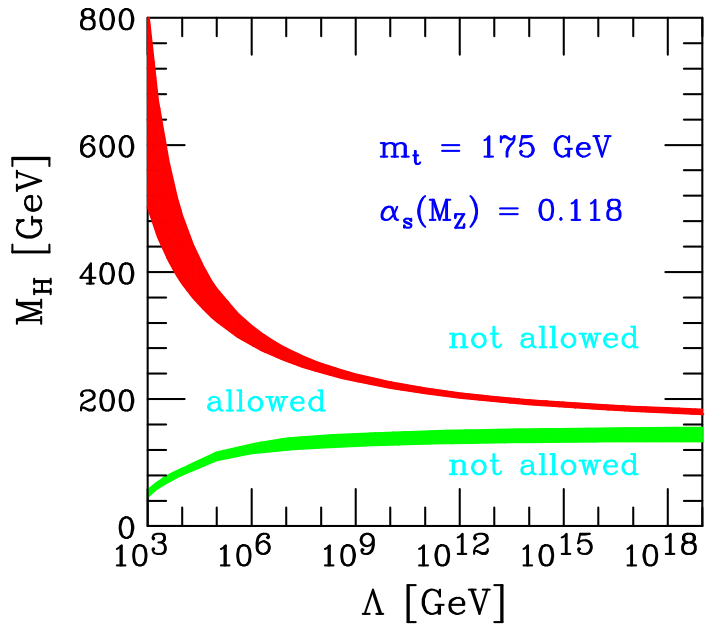,bbllx=18pt,
bblly=180pt,bburx=488pt,bbury=567pt,width=10cm}
\end{center}
\caption[fig1]
{The area between the two curves shows the 
allowed Higgs mass range assuming the validity of the 
Standard Model up to a scale $\Lambda$~\cite{mhiggssm2}.}
\end{figure}

The requirement of perturbative consistency of the theory up 
to a scale $\Lambda$ sets an upper bound on the SM Higgs mass, 
while arguments of vacuum stability suggest a lower Higgs
mass limit~\cite{mhiggssm2}, depending also strongly on the top mass. 
Taking the measured value of the top mass (m$_{t}$= 174.1$\pm$5 GeV)
and assuming that no new physics exists below the Planck scale, 
the Higgs mass should be around  $160 \pm 20$ GeV, as shown in figure 1.
\begin{figure}[htb]
\begin{center}
\epsfig{file=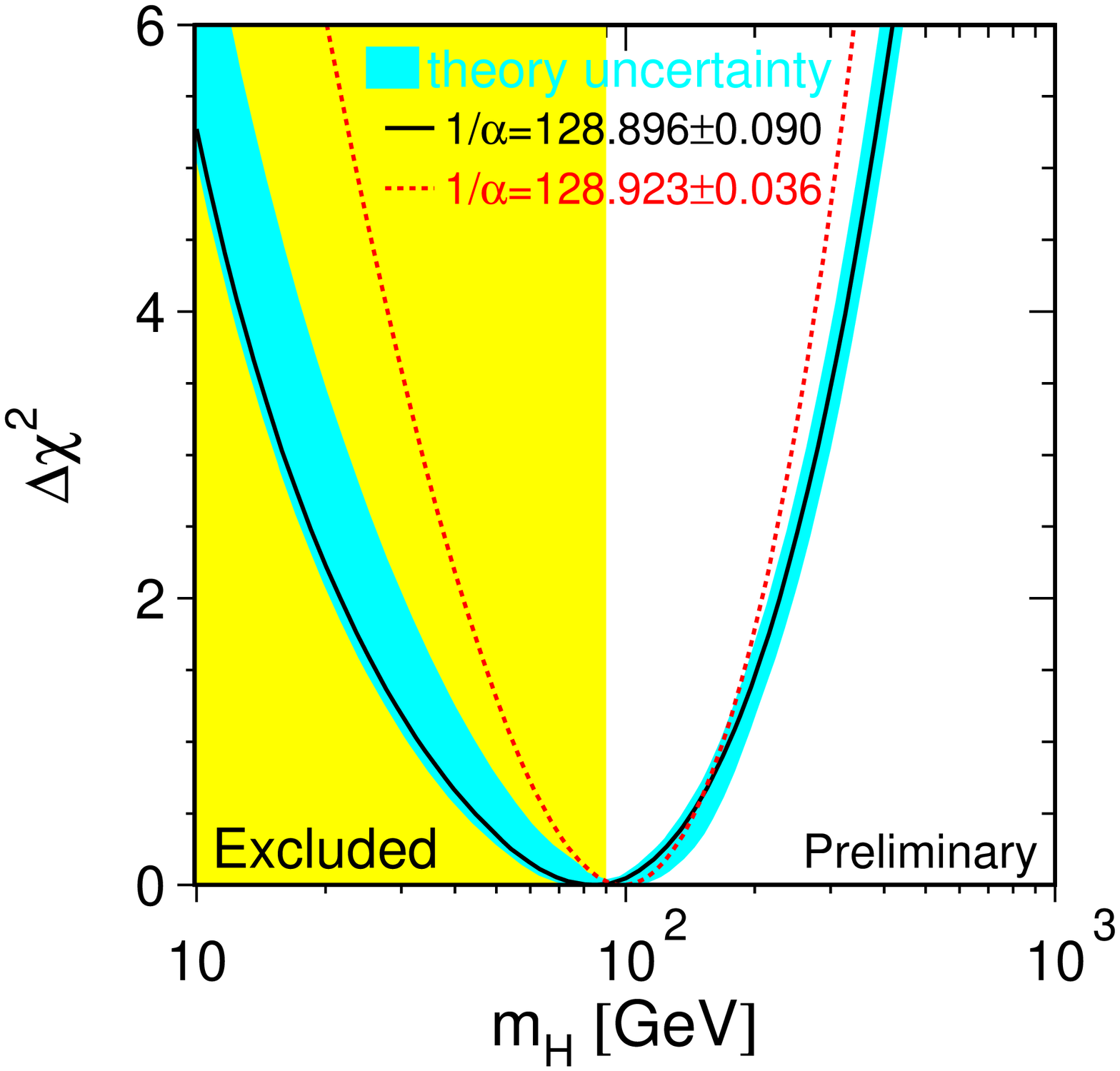,width=8cm}
\end{center}
\caption[fig2]{$\Delta \chi^{2}$ result of 
a fit to all electroweak observables assuming to have the 
Higgs mass as the only remaining free parameter~\cite{quast}.}
\end{figure}

Particle physicists have been searching for many years for 
the Higgs boson, from zero mass up to the highest masses 
accessible at existing particle accelerators.
At present, the four LEP experiments have ruled out the existence
of a Higgs with a mass of less than  95 GeV~\cite{LEPC98} .
From global fits to 
electroweak data one obtains an upper limit on the Higgs mass
of 280 GeV (95\% C.L.), as shown in figure 2~\cite{quast}.

By the end of the year 2000 one expects to discover at LEP200 a
Higgs boson up to $m_{H} \approx 106$ GeV, assuming 150 pb$^{-1}$ per 
experiment at $\sqrt{s}$ = 200 GeV. In case no Higgs signal is found, 
a mass limit of $\approx$ 109 GeV (95\% C.L.) will be placed~\cite{hlep2f}.

In the supersymmetric extension of the SM a set of new particles 
should exist with a mass scale around 1 TeV. 
The minimal version of the supersymmetric SM (MSSM) contains three 
neutral and two charged Higgs bosons. 
One of the neutral ones is expected to have a mass around 100 GeV 
and is therefore of particular interest for searches at LEP200.
So far the searches for this lightest Higgs boson 
resulted in a lower mass limit of about 80 GeV. 
The expectation for SUSY Higgs searches at LEP200 is 
summarised in figure 3~\cite{Daniel}. 
However, the discovery potential
at LEP200 depend strongly on the available energy. Thus few GeV increase in 
$\sqrt{s}$ could change our understanding of the Higgs sector dramatically.
\begin{figure}[htb]
\begin{center}
\epsfig{file=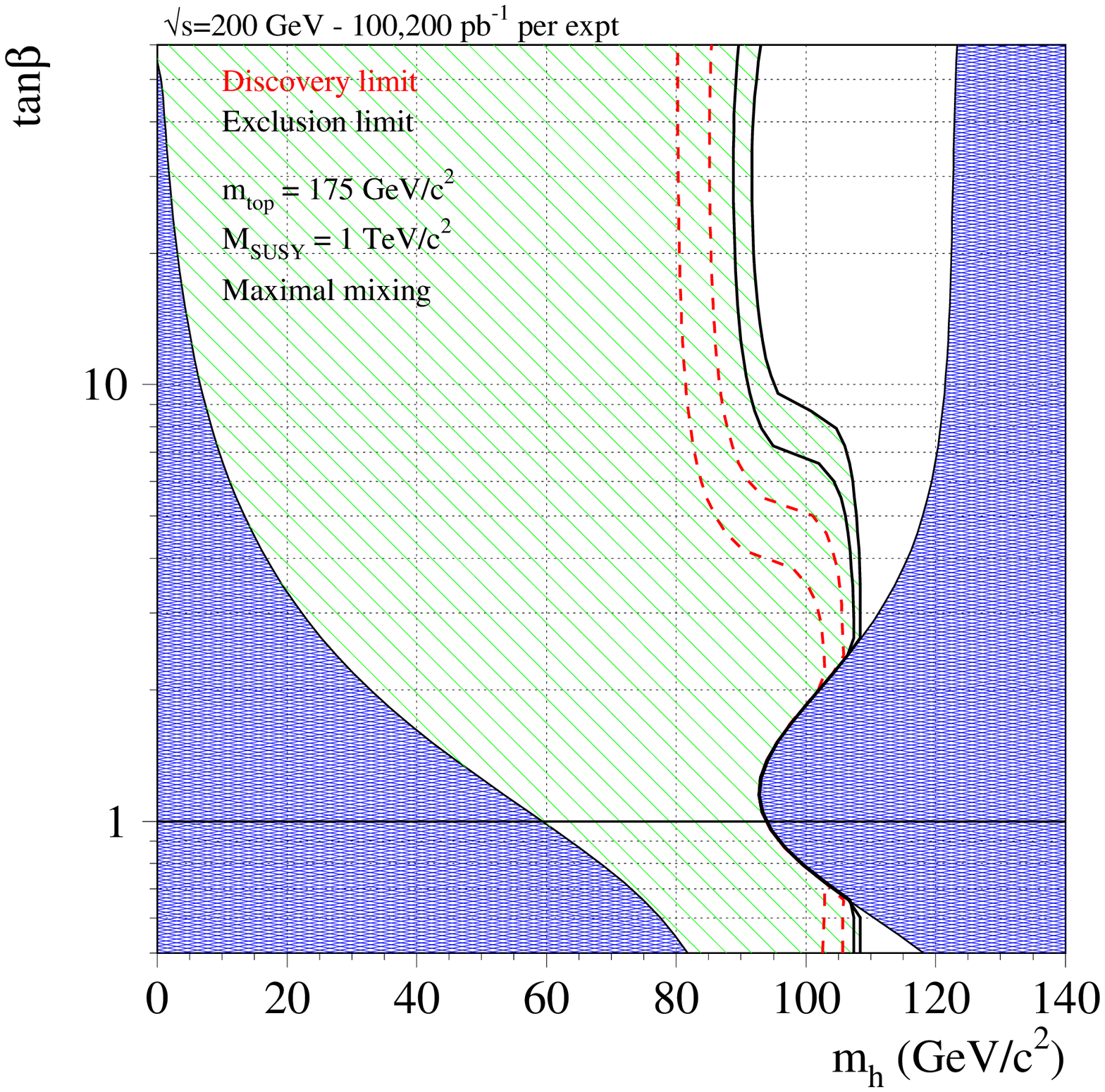,bbllx=19pt,bblly=145pt,bburx=554pt,bbury=672pt,width=6cm}
\epsfig{file=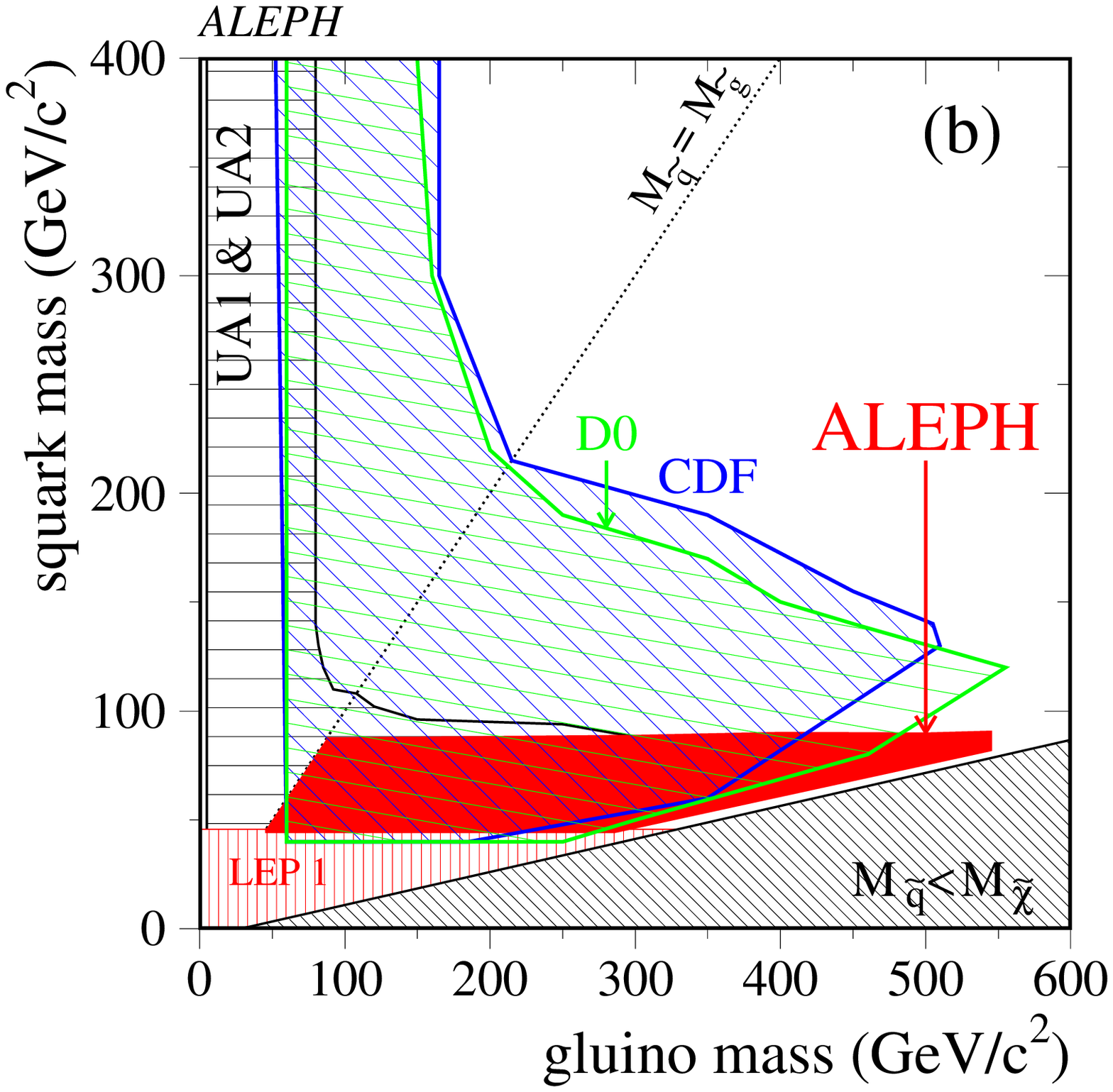,
bbllx=15pt,bblly=127pt,bburx=490pt,bbury=636pt,width=6cm}
\end{center}
\caption[fig3] {Expected sensitivity of SUSY Higgs searches (left figure)
at LEP200 ($\sqrt{s}$= 200 GeV) and recent lower mass limits for squarks and 
gluinos (right figure)~\cite{Daniel}.}
\end{figure}
%
\subsection {The sparticle sector}

Direct searches for sparticles at LEP200 have reached in most cases the 
kinematical limit, i.e. sparticle masses below $\approx$ 90 GeV are 
excluded~\cite{Daniel}. 
Searches for sparticles at the Tevatron have excluded gluino and squark masses
below about 250 GeV, as shown in figure 3.
With the data collected during RunII ($\approx$ 1fb$^{-1}$/year)
at the Tevatron, 
scheduled to start in 2000, one expects to reach 
gluino and squark masses of 300 to 400 GeV.
%
%
%

\section {The World of Physics at the LHC}
Discovering new phenomena in high energy physics experiments 
rely on the capability to separate {\em new}
from {\em known} phenomena . The methods used exploit the 
different kinematics of signals and backgrounds 
in searching for new mass peaks, or comparing
p$_{T}$ spectra of leptons, photons and jets and their 
angular correlations with SM predictions. Other searches exploit 
the missing transverse energy  signature which 
might originate from neutrinos or 
neutrino--like objects, or simply 
from detector imperfections. Depending on the particular physics process,
different aspects of the detector performance parameters are important. 
The search for mass peaks requires in general excellent 
energy and momentum resolution for individual particles. 
Searches based on the missing transverse energy signature  
require detectors with hermetic calorimeter coverage up to $|\eta|$= 5.  

Figure 4 shows the world of physics to be explored with multi--TeV 
proton-proton collisions at the LHC. This world is divided into 
sectors according to the detector requirements for measuring photons, 
leptons (e, $\mu, \tau$), missing transverse energy, jets and the 
capability to identify b--jets. With the expected LHC detector 
performance as described in the following sections, 
the SM Higgs mass range can be covered from the expected LEP200 limit 
all the way up to about 1 TeV. Entering the world of Supersymmetry, 
the same detector performance  allows to cover, to a large extent, the 
different  SUSY signatures. The world of new heavy vector bosons,
as for example predicted by the strongly interacting Higgs sector, can be 
explored with an excellent lepton resolution.

Although the most exciting discoveries will be those of totally 
unexpected new particles or phenomena, one can only
demonstrate the discovery potential of the proposed experiments 
using predicted new particles. However, the experiments 
designed under these considerations should also allow the 
discovery of whatever new phenomena might occur in 
multi--TeV pp collisions.
\begin{figure}[htb]
\begin{center}
\rotatebox{-90}{
\epsfig{figure=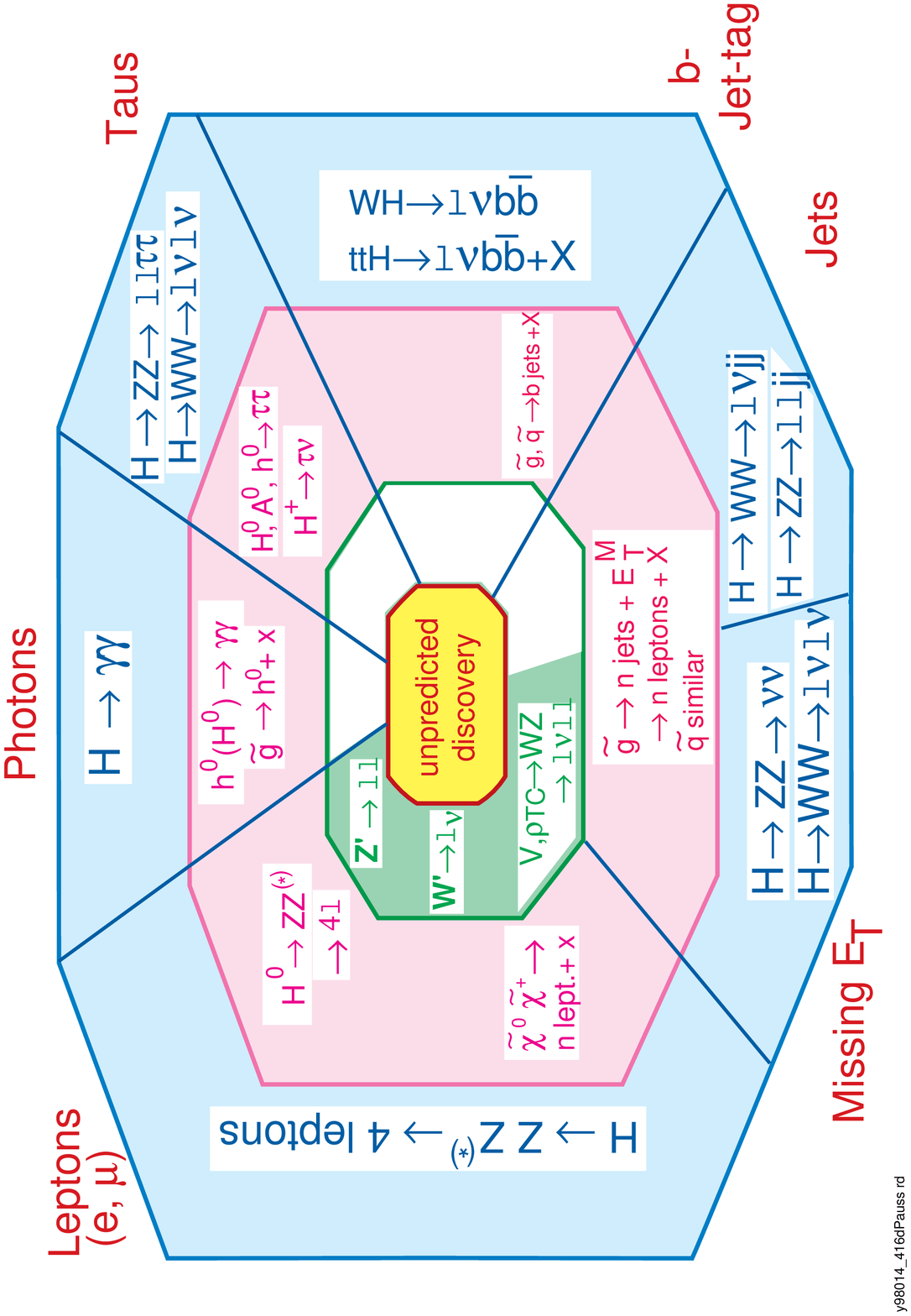,
width=11.cm,height=15.cm}}
\end{center}
\caption[fig4]{The world of physics at the LHC}  
\end{figure}

%
%
%
\subsection {The Experimental Challenge at the LHC}
The total cross--section at hadron colliders is very large, i.e. about 100 mb
at the LHC, resulting in an interaction rate of $\approx$ 10$^{9}$ Hz 
at the design luminosity. Figure 5 shows the expected energy dependence
of the total cross section and of some interesting physics processes
which have much smaller cross sections.
The detection of processes with signal to 
total cross--section ratios of about 10$^{-12}$, as for example for a 100 GeV 
Higgs decaying into two photons, will be a difficult experimental challenge. 

\begin{figure}
\begin{center}
\includegraphics*[scale=0.8,bb=0 60 600 780]
{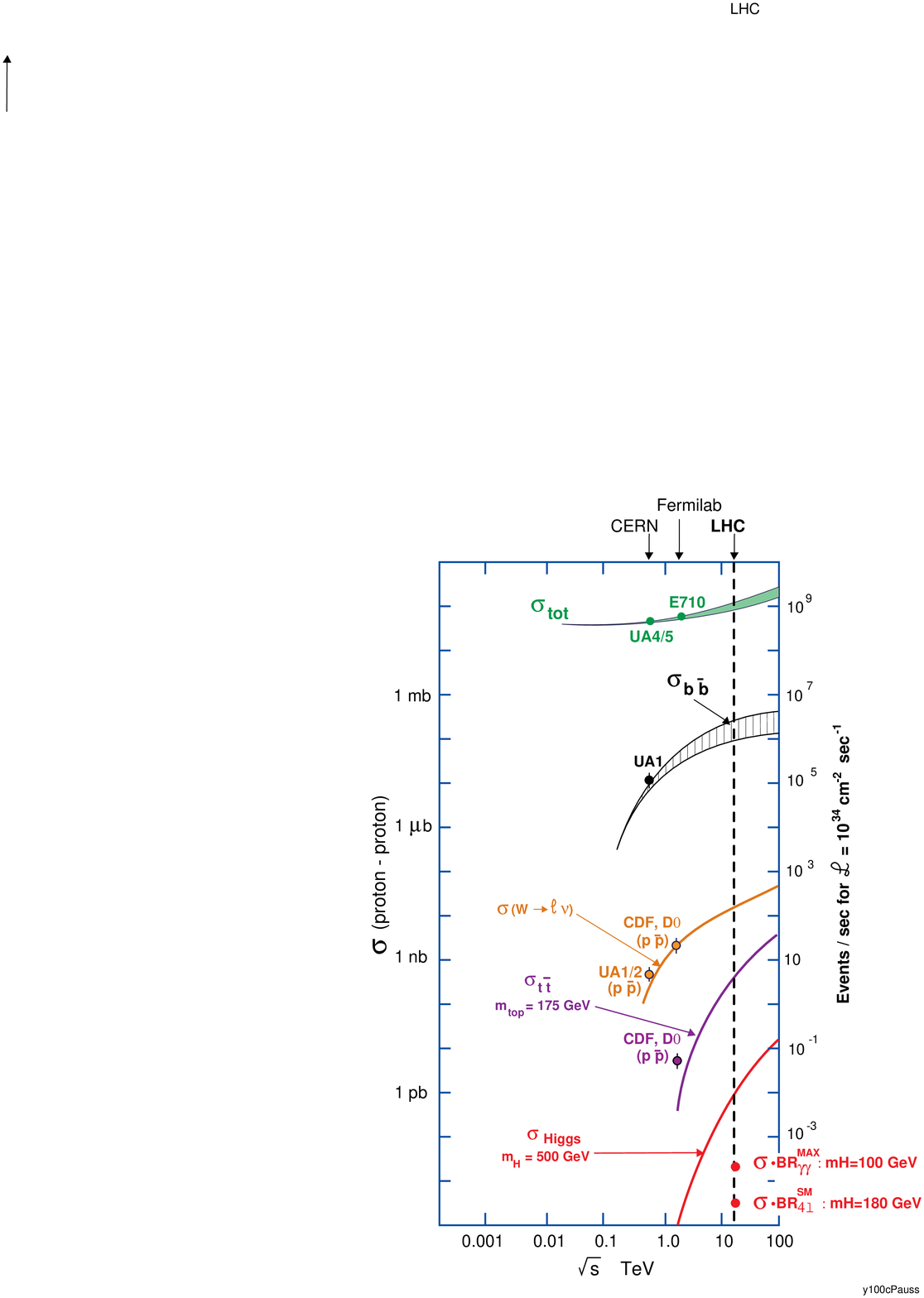}
\end{center}
\caption[fig5]{Energy dependence of some characteristic 
cross--sections at hadron colliders.}
\end{figure}


Many of the above mentioned new particles decay into W and Z bosons, 
charged leptons or photons. Ws and Zs will have to be detected 
through their leptonic decays because hadronic decay modes will 
be overwhelmed by the QCD~\cite{james}  background. 
These purely leptonic modes 
lead to very small branching fractions. 
In order to observe such signals, a machine with high constituent 
centre--of--mass energy and  high luminosity is required.

The LHC fulfils these requirements, but the high luminosity leads to 
difficult experimental conditions: with an inter--bunch crossing time 
of 25 ns at design luminosity, on average  20  
interactions (``minimum bias events'') are expected per crossing, 
resulting in about 1000 charged tracks every 25 ns, in the 
pseudorapidity range of  $|\eta| \leq $ 3. 
Therefore, at peak luminosity, on average  2.2 charged particles 
are expected every 25 ns in a 2$\times$2cm$^{2}$ cell 
at a distance of 7.5 cm from the 
interaction point at $\eta$= 0. This example shows that the 
inner tracking detectors have to operate in a hostile environment. 
Such high particle fluxes will make track reconstruction difficult. 
Simulation results make us believe that a very large number of
electronic channels and good time resolution should 
nevertheless guarantee a high track--finding efficiency.

The expected 10$^{9}$ inelastic pp events 
per second at design luminosity will also 
generate a hostile radiation environment. 
This results in high radiation levels (high integrated dose) 
and in a large flux of low energy neutrons in the experimental area. 
As an example, figure 6 shows the radiation environment in CMS~\cite{ecaltdr}. 
Radiation hard detectors and electronics are therefore required. 
Induced activity in the forward calorimeters has to be taken into 
account for long--term access and maintenance.

\begin{figure}[htb]
\begin{center}
\includegraphics*[scale=0.6,bb=8 22 527 529]{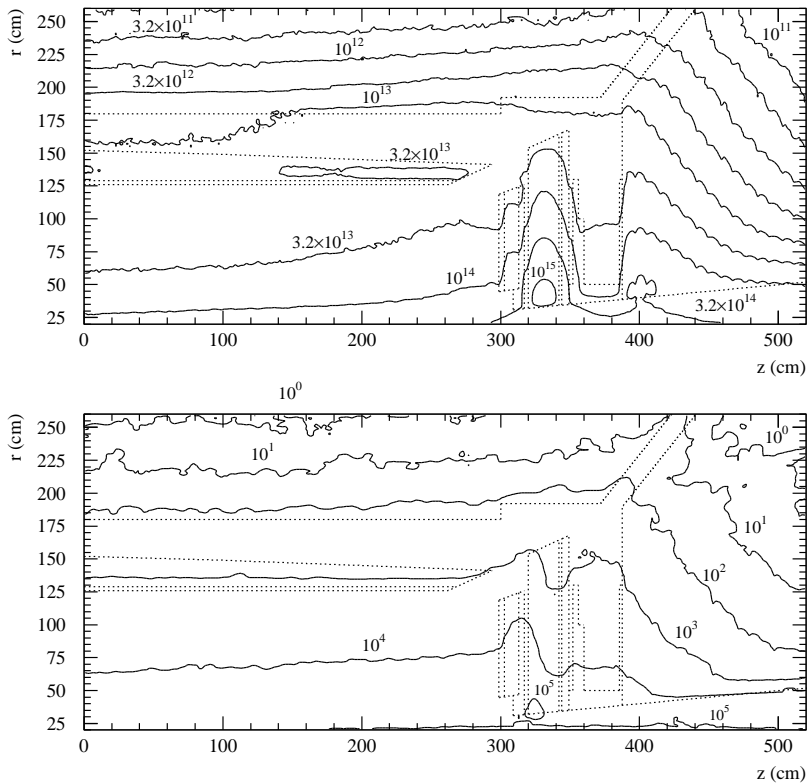}
\end{center}
\caption[fig6]{Fluence of neutrons and charged hadrons in cm$^{-2}$ 
(upper plot) and radiation dose  in Gy (lower plot)  in the CMS 
calorimeter region. Values correspond to an integrated 
luminosity of 5$\times$10$^{5}$ pb$^{-1}$~\cite{ecaltdr}.}
\end{figure}
%
%
%
%

\section {Design Objectives of ATLAS and CMS}

An important aspect of the overall detector design is the magnetic
field configuration. Large bending power is required to 
measure precisely high--momentum muons and other charged particles. 
The choice of the magnet structure strongly influences the remaining 
detector design. A {\em solenoid} provides bending in the transverse 
plane and thus facilitates the task of triggering on muons, 
which are pointing to the event vertex, so that one can take 
advantage of the small transverse dimensions of the 
beam ($ 20\mu$m). A drawback of a solenoid with limited length is the 
degradation of momentum resolution in the forward direction; therefore
either a very long solenoid is required or an endcap toroid 
system has to be added. 
The main advantage of a {\em toroid} is a constant p$_{T}$ resolution 
over a wide rapidity range. However, the closed configuration 
of a toroid does not provide magnetic field for the inner tracking, 
thus an additional solenoid is required to measure 
the momenta of charged tracks in the inner tracking detectors.

The identification and precise measurement of electrons, photons 
and muons over a large energy range, complemented by 
measurements of jets and missing transverse energy are the basic 
design goals of the  ATLAS  and 
CMS detectors. In addition, a good impact--parameter 
resolution and secondary vertex reconstruction will be 
important for b-tagging.

The ATLAS~\cite{atlastp} collaboration has chosen a magnet configuration based 
on a superconducting air--core toroid, complemented by a 
superconducting solenoid of 2 T around the inner tracking 
detectors. The thin solenoid is followed by 
a high--granularity liquid--argon sampling calorimeter. 
In the toroidal magnet configuration the muon triggering, 
identification and precision measurement
can be entirely performed in the muon spectrometer, 
without using the inner detectors.
%
%
\begin{figure}[htb]
\vspace{3.5cm}
\begin{center}
\epsfig{file=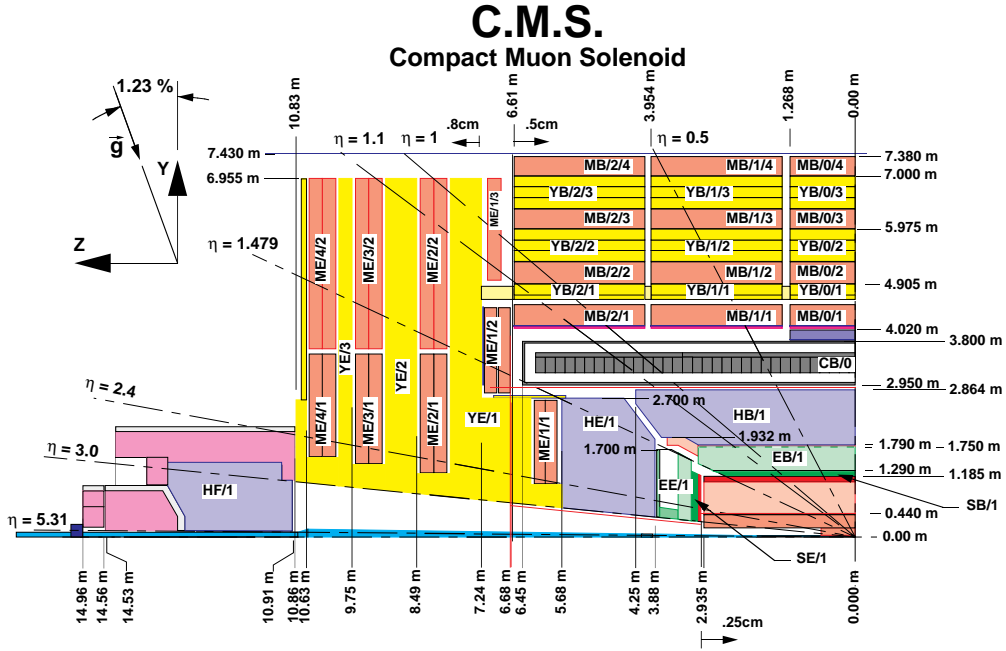,width=10.cm}
\end{center}
\caption[fig7]{Schematic 1/4 view of the CMS detector. The 
total weight amounts to about 12500 tons.}
\end{figure}

The CMS~\cite{cmstp} detector will use a high--field superconducting 
solenoid (4 T) allowing for a compact design of the muon 
spectrometer (see figure 7). The inner coil radius of about 3 m 
is large enough to accommodate the inner tracking system and the 
calorimeters. For the electromagnetic calorimeter 
PbWO$_{4}$ crystals have been
chosen. The hadron calorimeter (also located before the coil) 
consists of copper absorber plates and scintillator tiles. 
Muons are triggered, identified and measured in four identical muon stations 
inserted in the return yoke. Their momenta are measured 
independently in the inner tracking chambers to improve the overall 
momentum measurement.

%
%
%
%
%

\section{Subdetector Requirements and Performance Figures}
In the following we discuss the different subdetector requirements
and illustrate the expected performance. The performance 
figures are obtained from
detailed simulation studies, and wherever possible, 
they are based on the actually measured 
performance of prototype detectors obtained from various R\&D projects.
\subsection{The muon system}
The muon system has to fulfil three basic tasks: (i) identification,
(ii) momentum measurement and (iii)
triggering. The latter is very  challenging in hadron collider experiments.
Muons have the advantage that they
can be identified inside jets and can therefore be used 
for  b--tagging (b $\rightarrow$ $\mu$ + X) as well as for 
measuring the energy flow around leptons inside  jets and 
thus evaluating  the  efficiency
of isolation cuts. Furthermore,
the possibility to trigger on and identify muons down to low p$_{T}$ 
increases the acceptance for important physics processes 
(e.g. H $\rightarrow$ ZZ$^{*}$ $\rightarrow$ 4$\mu$ and CP violation studies).

The performance of the $\mu$--system is determined by (i)
pattern recognition:
hits from $\mu$--tracks  can be spoilt by 
correlated background ( $\delta$'s, electromagnetic showers, punch-through) 
and uncorrelated background ( neutrons and associated $\gamma$s), 
(ii) momentum resolution:  many factors 
influence the muon resolution, like for example,
 multiple scattering,
fluctuations in the energy loss, 
accuracy of tracking devices in the muon spectrometer,
alignment and the magnetic field map and
(iii)  1$^{st}$-- level trigger:
the $\mu$--rate is dominated by $\pi$ and $K$ decays up to 4 GeV 
and by b-- and  c--quark decays from 4 to 25 GeV. 
Figure 8 shows the $\mu$--trigger rate at the design 
luminosity. At p$^\mu _{T}$ = 10 GeV a  1$^{st}$--level trigger 
rate of about 10$^{4}$ Hz is expected~\cite{cmsmutrig}. 
The trigger rate has to be 
adjustable by moving the threshold in a wide range of 
p$^{\mu}_{T}$ to account for different background conditions
without however loosing efficiency for important physics channels.
\begin{figure}[htb]
\begin{center}
\mbox{
\epsfig{file=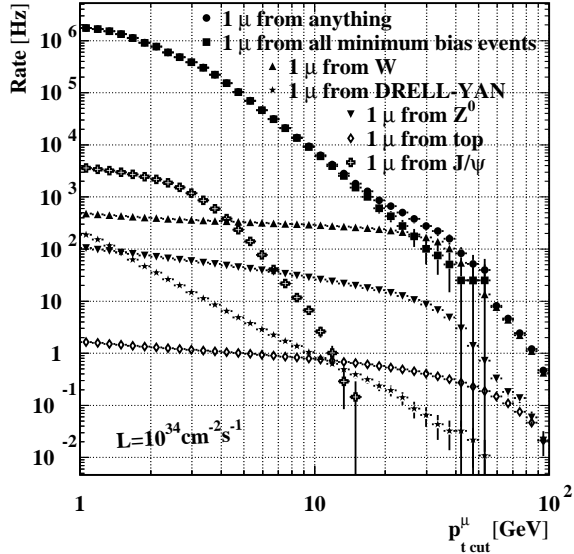,width=8.0cm}
}
\end{center}
\caption[fig8]{Predicted CMS muon trigger rates at the 
LHC design luminosity~\cite{cmsmutrig}.}
\end{figure}

The ATLAS superconducting air--core toroid system is 
optimised for stand--alone measurements~\cite{atlasmu}. 
The $\mu$--spectrometer consists of precision chambers 
(monitored drift tubes and cathode strip chambers) which 
require a high--accuracy tracking (50$\mu$m/chamber), 
to be aligned to 30$\mu$m. 
Resistive plate chambers and thin gap chambers are used for triggering.
The expected resolution ranges from about 2-3\% at p$^{\mu} _{T}$ = 100 GeV
to about 11\% at  p$^{\mu} _{T}$ = 1 TeV.

The CMS $\mu$--system consists of 4 identical 
muon stations~\cite{cmsmu}. In the barrel, 
12 layers of drift tubes are used with $\sigma$ = 250 $\mu$m per layer. 
Cathode strip chambers (6 layers) with $\sigma$ = 75--150 $\mu$m are 
implemented in the endcaps. 
The chambers have to be aligned to 100--200 $\mu$m. 
The favourable aspect ratio (length/radius) of the superconducting coil 
allows good $\mu$--momentum resolution up to pseudorapidity of 2.5 
with a single magnet. The 4 T field provides a powerful combined 
measurement up to $|\eta | \approx$ 2.5. 
Figure 9 shows the expected momentum resolution in the CMS experiment
and illustrates the resulting experimental $4\mu$--resolution 
as a function of m$_{H}$. This resolution is compared to the width of 
the SM--Higgs and the expected $4e$--resolution. 
The combined experimental $4l$--mass resolution dominates the measured 
width up to m$_{H} \approx$ 250 GeV.
\begin{figure}[htb]
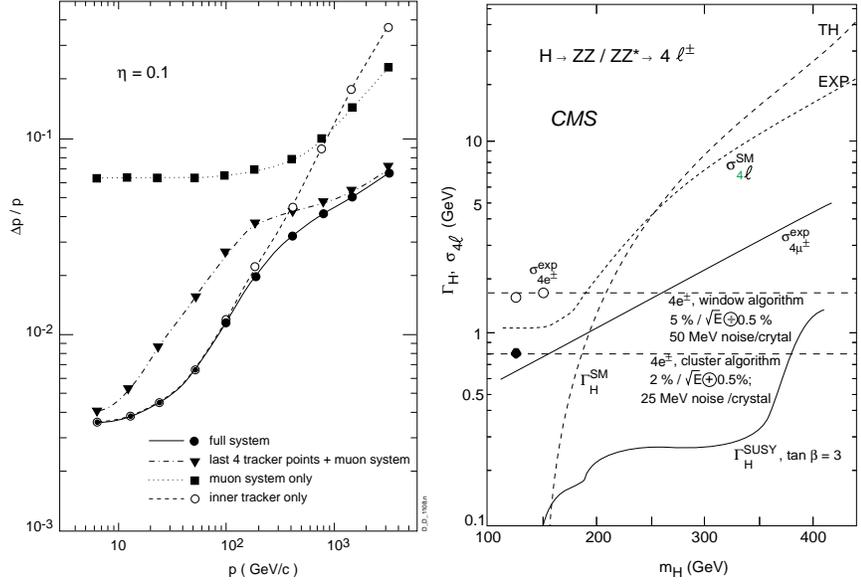

\begin{center}
\includegraphics*[scale=0.3,bb=9 7 540 745]{D_Denegri_1108n.ill}
\includegraphics*[scale=0.32,bb=33 34 569 713]{D_Denegri_0876n.ill}
\end{center}
\caption[fig9]{Expected muon momentum resolution 
in the CMS experiment (left) and the resulting experimental 
$H \rightarrow 4\mu$ mass resolution (right) compared with the 
expected width of the Higgs boson~\cite{cmsmu}.}
\end{figure}
%
%

\subsection{Hadron calorimeter}

The performance of the hadron calorimeter can be characterised by 
the jet--jet mass resolution and the missing transverse 
energy (E$^{miss}_{T}$) resolution.
The mass resolution depends on the calorimeter resolution but also
on the jet algorithm, 
fragmentation, energy pile--up and the cone--size for jet reconstruction.
Studies of the effect on the mass resolution for W,Z $\rightarrow jet+jet$, 
using the high--mass Higgs decay  
H $\rightarrow$ \  ZZ(WW) $\rightarrow$ $lljj$ have shown that a 
calorimeter granularity of  $\Delta\eta \times \Delta\phi \approx$
0.1$\times$0.1 is sufficient to measure jets from a boosted W or Z 
with high efficiency.
A good E$_{T}^{miss}$ resolution requires a hermetic
calorimeter coverage up to $|\eta|$=5 (i.e. cracks and dead areas 
have to be minimized). This requirement has been taken into 
account in the design of the ATLAS~\cite{atlashad}, and CMS~\cite{cmshad} 
calorimeter system.
Figure 10 illustrates the jet resolution for E$^{jet}_{T}$ = 100 GeV
and the expected E$_{T}^{miss}$ resolution in ATLAS at 
low luminosity.

\begin{figure}[htb]
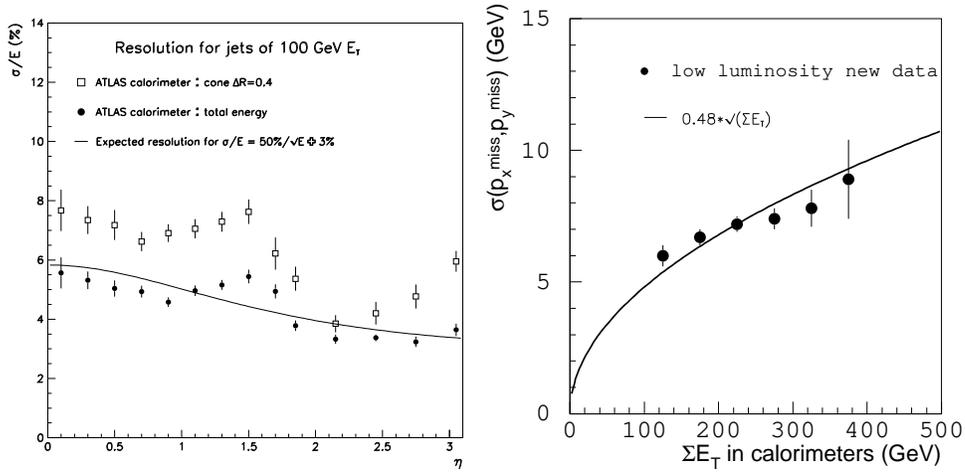

\begin{center}
\mbox{
\epsfig{file=AJETRES.epsi,height=6cm}
}
\epsfig{file=AETMISS.epsi,height=6.2cm}
\end{center}
\caption[fig10] {The expected energy resolution
for jets with E$^{jet}_{T}$ = 100 GeV (left) and 
the missing transverse energy resolution (right) in ATLAS~\cite{atlasperf}.}
\end{figure}
%

%
%

The ATLAS barrel hadron calorimeter consists of 
iron with scintillating tiles. The endcaps use a parallel plate 
design for the copper liquid--argon calorimeter. 
The forward calorimeter is made of tungsten liquid--argon with very small 
gaps (0.25 to 0.5 m) to limit the ion build--up.

The barrel and endcaps hadron calorimeter in CMS is placed 
inside a 4 T field.
CMS has chosen a  copper/plastic--scintillator
system.
In the forward direction, due to the
high radiation environment (absorbed dose $\sim$ MGy/year,  
neutron flux $\sim$10$^{9}$cm$^{-2}$s$^{-1}$),
an iron/quartz fibre system was selected.
More details about the proposed hadron calorimeters 
can be found in reference~\cite{jim}.
%
\subsection{Electromagnetic calorimeter}

The performance of the electromagnetic calorimeter
is best demonstrated using the  
H $\rightarrow$ $\gamma\gamma$ reaction. For m$_{H}$ = 100 GeV the 
Higgs width is only a few MeV, therefore the measured mass 
resolution is entirely dominated by the experimental resolution.

The CMS design goal requires a high resolution electromagnetic calorimeter,
therefore a fully active (homogeneous) calorimeter 
consisting of PbWO$_{4}$ crystals has been chosen~\cite{ecaltdr}. 
%
%
The photon resolution 
expected in CMS is summarised in the following table: \\

\begin{center}
\begin{tabular}{l|c|c}
\hline
\multicolumn{3}{|c|}{Photon energy resolution in CMS} \\
\hline  \hline 
  & Barrel ($\eta$=0) & Endcap ($\eta$ =2)  \\ \hline
   stochastic term & 2.7\%/$\sqrt{E}$ & 5.7\%/$\sqrt{E}$    \\
   constant term & 0.55\% &0.55\%    \\
   E$_{T}$ noise (L=10$^{33}$cm$^{-2}$s$^{-1}$) & 155 MeV & 205 MeV    \\
   E$_{T}$ noise (L=10$^{34}$cm$^{-2}$s$^{-1}$) & 210 MeV & 245 MeV    \\ 
\hline\hline
\end{tabular}
\end{center}

ATLAS has chosen a sampling calorimeter which provides 
longitudinal and transverse segmentation~\cite{atlashad}. An accordion 
structure 
in liquid argon with lead absorbers is used.
The expected $\gamma$--resolution as function of pseudorapidity 
for E$^{\gamma} _{T}$ = 50 GeV is shown in figure 11.
More details about the proposed electromagnetic calorimeters 
can be found in reference~\cite{jim}.
%
\begin{figure}[htb]
\begin{center}
\mbox{
\epsfig{file=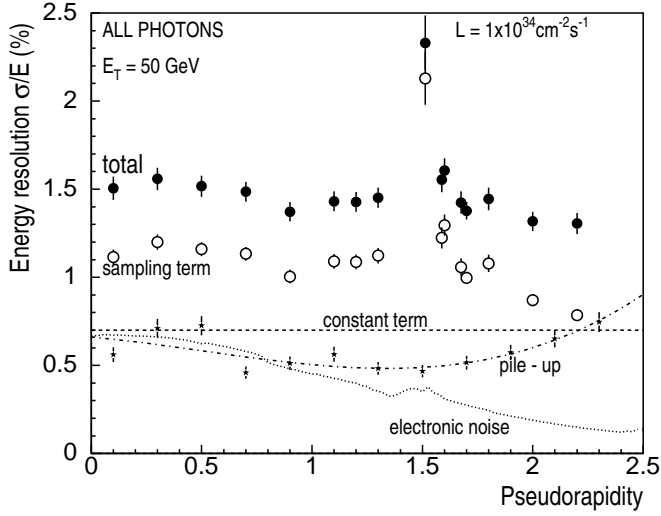,height=7.cm}
}
\end{center}
\caption[fig11]{ $\gamma$--resolution as function of pseudorapidity 
for E$^{\gamma} _{T}$ = 50 GeV in the ATLAS detector~\cite{atlasperf}. }
\end{figure}
%
%
%
\subsection{Inner tracking detectors}

The inner tracking system provides precise momentum 
and impact parameter and secondary vertex 
measurements for charged particles. 
It is also essential for $e$ and $\tau$ identification,
and the calibration of the electromagnetic calorimeter 
with electrons, using the p/E matching.

Silicon detectors are already  very successfully used in HEP 
experiments for vertex determination. However, a transition 
from detector sizes of $\mathcal{O}(m^{2})$
with electronics at the periphery of the sensitive area to sizes of
$\mathcal{O}(50 m^{2})$ with 
electronics distributed over the sensitive area is needed.
Furthermore, 
tracking at the LHC must cope with high instantaneous and integrated rates.
The tracking system  must operate at an integrated dose
(500 fb$^{-1}$) of Mrad to 30 Mrad and a neutron fluence of 10$^{14}$
to 10$^{15}$ n/cm${^2}$, therefore cooling is required.
At the LHC silicon detectors are used for r $\leq$  60 cm and  
gaseous detectors (microstrip gas chambers (MSGC) and
transition radiation detectors (TRD)) are used for $\geq$ 60 cm.
More details about silicon detectors can be found in 
reference~\cite{iris}.
%

The design goals for the tracking system are:
\begin{itemize}
\item for isolated leptons in the CMS detector:
$\Delta p_{T}/p_{T} \sim$ 0.1 p$_{T}$ (TeV). \\
in ATLAS: $\Delta p_{T}/p_{T} \leq$ 30\% at 500 GeV, 1--2\% at 20 GeV  
\item high--p$_{T}$ track reconstruction efficiency: for isolated tracks:
$\varepsilon >$95$\%$, and within jets: $\varepsilon >$90$\%$
(ghost tracks $<$ 1$\%$ for isolated tracks).
\item impact parameter resolution: at high p$_{T}$ 
$\sim$20$\mu$m (r$\phi$), $\sim$100$\mu$m (z)
\end{itemize}
%
%

The ATLAS tracking system~\cite{atlastracker} is located inside 
a solenoidal field of 2 T and consists of
3 pixel layers ( $\sigma _{r\varphi} \sim $ 12$\mu$m),
4 silicon strip layers ($\sigma _{r\varphi}  \sim $ 16$\mu$m) and
$\sim$ 40 transition radiation tracking (TRT) layers 
( $\sigma _{r\varphi} \sim $ 170$\mu$m). 

The CMS tracking~\cite{cmstracker} is inside a 4 T field and consists of
2 pixel layers ($\sigma _{r\varphi}$, $\sigma _{z} \sim $ 15$\mu$m),
5 silicon strip layers ($\sigma _{r\varphi}  \sim $ 15$\mu$m) and 
6 MSGC layers ($\sigma _{r\varphi} \sim $ 50$\mu$m).

The pattern recognition and momentum resolution is effected by 
conversion and bremsstrahlung. A low material budget is desirable 
also in order to maintain the electromagnetic calorimeter performance. 
The ATLAS and CMS material budget is plotted in 
figure 12 and shows that the  inclusion of support structures and cables 
increases the material budget beyond the desirable value.
\begin{figure}[htb]
\begin{center}
\epsfig{file=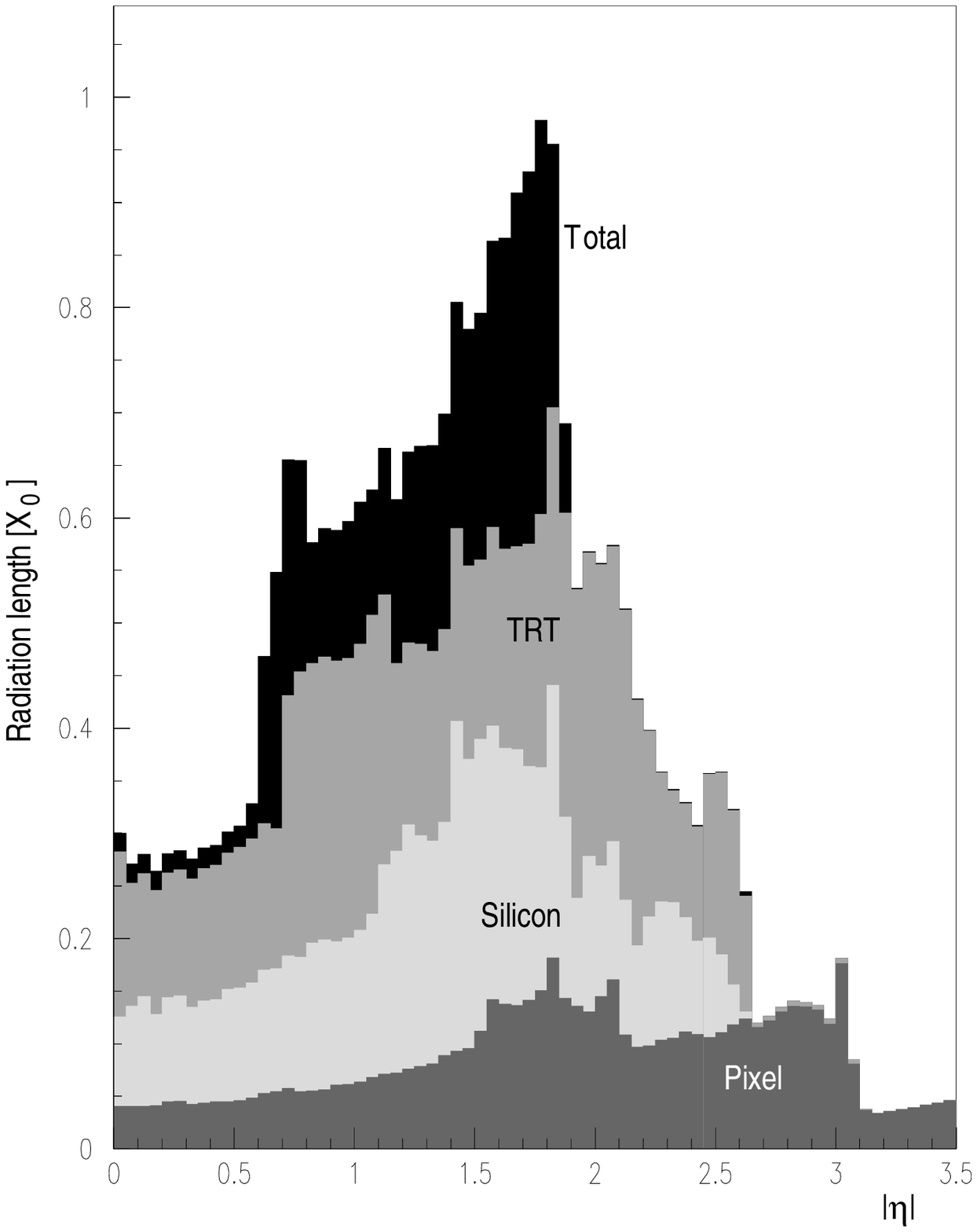,height=5.cm,clip=}
\epsfig{file=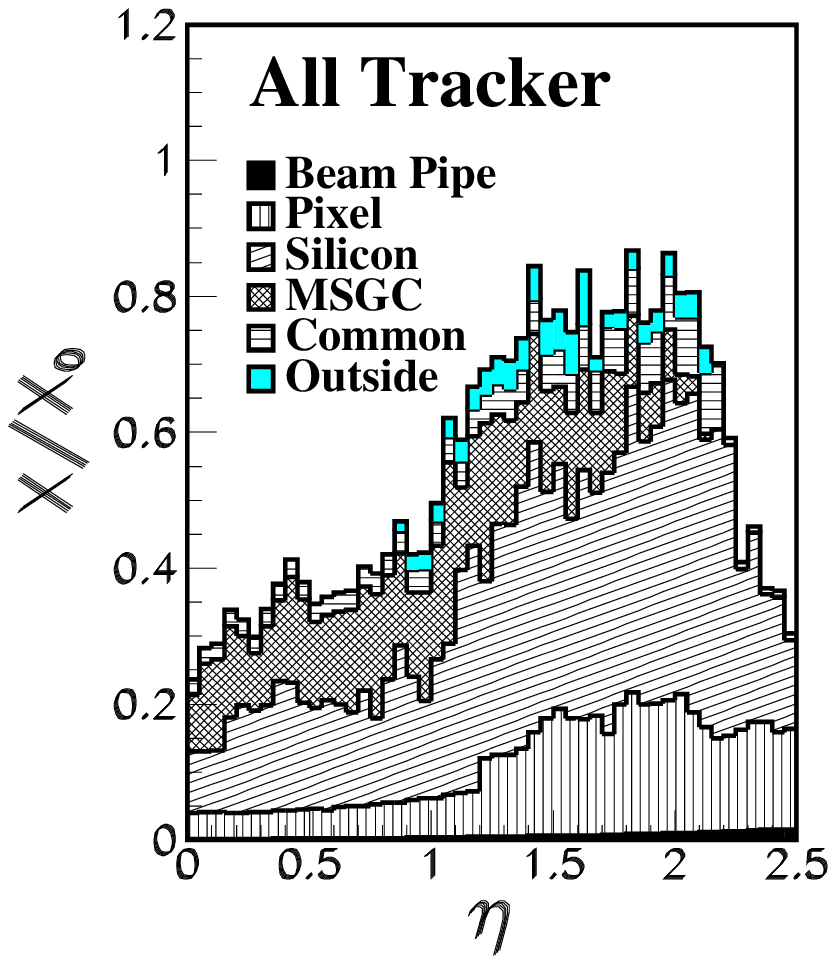,height=5.5cm,clip=}
\end{center}
\caption[fig12]{Material budget of the inner tracking system 
in ATLAS (left)~\cite{atlastracker} and in CMS (right)~\cite{cmstracker}.}
\end{figure}
Figure 13a and b show the 
expected p$_{T}$ resolution as a function of $\eta$ 
in ATLAS and 
the impact parameter resolution in CMS as a function of p$_{T}$ for
different $\eta$--values.
\begin{figure}[htb]
\begin{center}
\epsfig{file=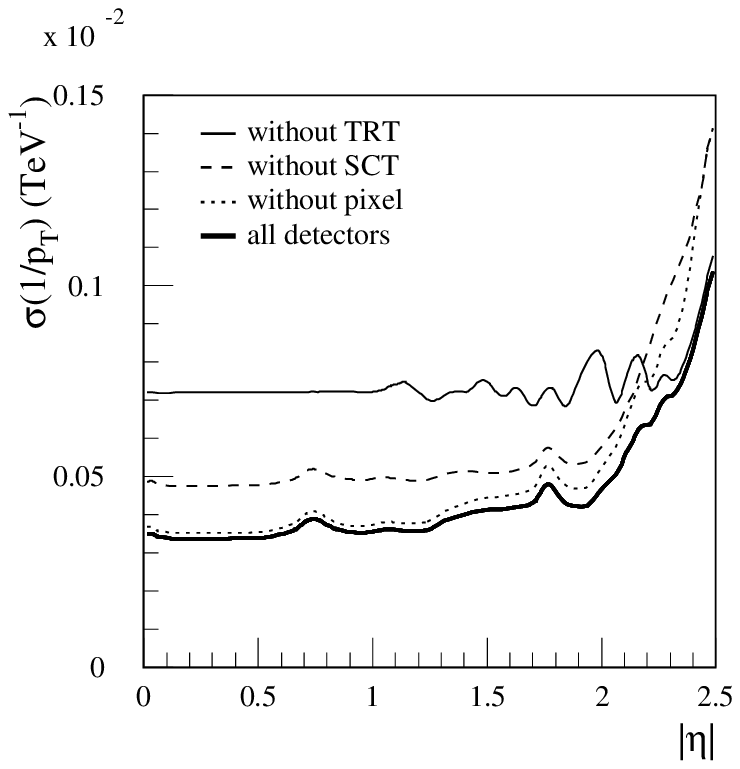,height=5.5cm}
\includegraphics*[scale=0.42,bb=110 183 494 561]{D_Denegri_0657n.ill}
\end{center}
\caption[fig13]{The expected p$_{T}$ resolution 
for high--p$_{T}$ tracks in
the ATLAS tracking system (left). The effect of removing complete 
subdetectors is also shown~\cite{atlastracker}. \newline
The figure on the right shows the impact parameter resolution 
in CMS as a function of 
p$_{T}$ for three different $\eta$--values~\cite{cmstracker}. }
\end{figure}
%
%
%
%

\subsection{Data acquisition and trigger}

The challenge at the LHC will be the reduction of 10$^{9}$ Hz interaction 
rate to about 100 Hz  output rate on tape for further off--line analysis. 
The on--line data reduction will 
proceed via different trigger levels. At the first level, 
local pattern recognition and energy evaluation on prompt 
macro--granular information will provide particle identification 
such as high--p$_{T}$  electrons, muons and missing E$_{T}$. 
Level--1 will select events at 10$^{5}$ Hz.
For level--2, finer granularity and more precise measurements will be 
used together with event kinematics and topology. 
By matching different subdetectors, clean particle signatures should be 
selected (e.g. W, Z, etc.), resulting in a level--2  rate of 10$^{3}$ Hz. 
Finally, event reconstruction and on--line analysis will result in physics 
process identification, leading to an output rate for further 
off--line analysis of about 100 Hz.
\begin{figure}[htb] 
\begin{center}
\epsfig{figure=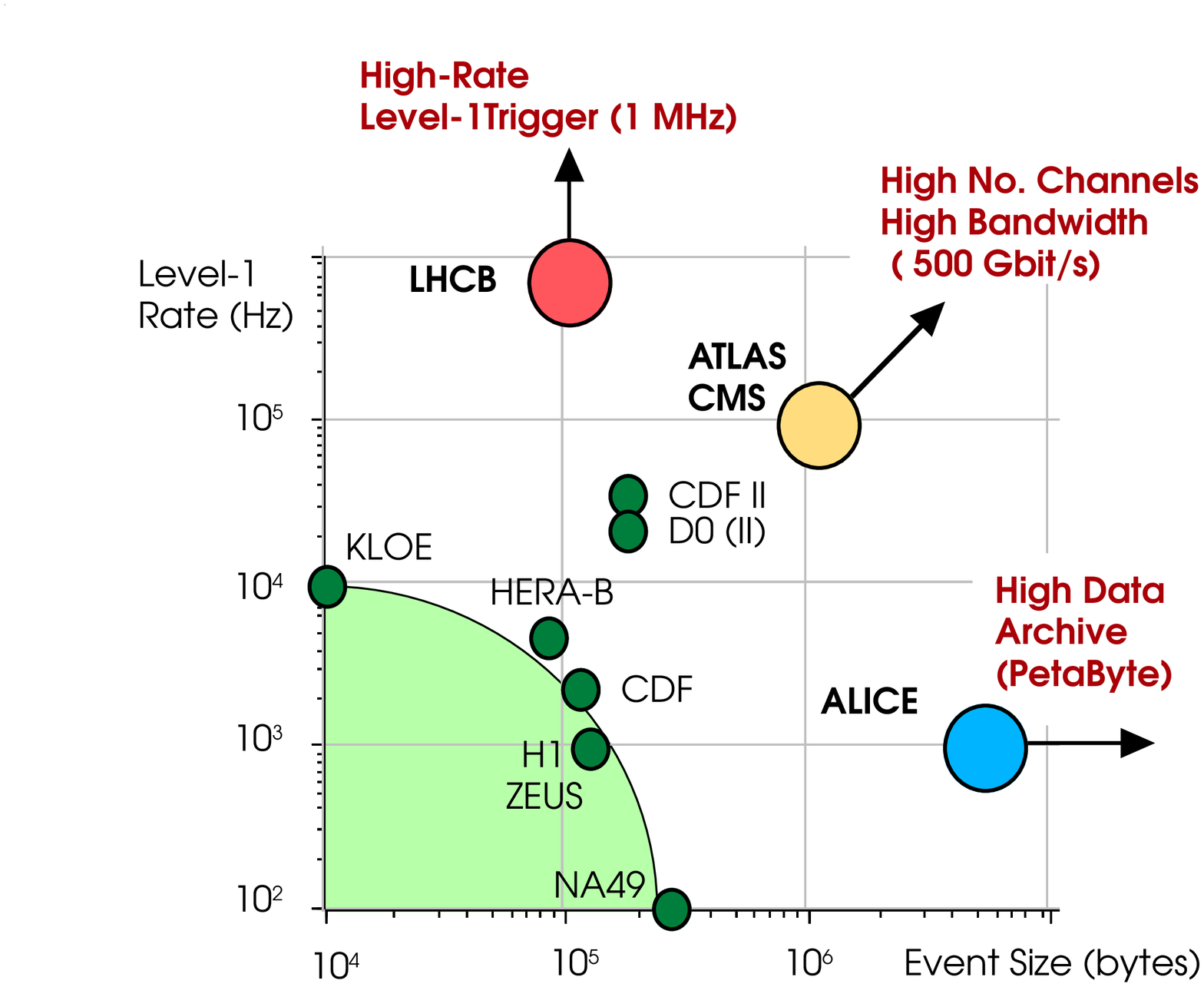,
width=10.0cm}
\end{center}
\caption[fig14]{Expected evolution for trigger and data acquisition
from existing 
experiments to future experiments at the LHC~\cite{paris}. }
\end{figure}
Figure 14 shows the trigger/DAQ 
evolution in terms of level-1 
rate and event size, from existing experiments to 
experiments at the LHC~\cite{paris}.
To illustrate the very demanding requirements for the trigger/DAQ system,
we recall some interesting numbers:
\begin{itemize}
\item the CERN (Lab--wide) computing power available in 1980 
was comparable to that of a modern desktop computer (1995),
\item the total number of processors in the LHC event filter 
equals the number of workstations and personal computers 
running at CERN in 1995 (4000),
\item during one second of LHC running, the data volume 
transmitted through the readout network is equivalent to the 
amount of data moved in {\em one day} by CERN network system 
(FDDI, Ethernet, local nets) in 1995,
\item the 'slow-control' data rate of an LHC experiment 
(temperature, voltage, status, etc.) is comparable to the 
current LEP experiment readout rate (100kByte/s),
\item the data rate handled by the LHC event builders 
(500 Gbit/s) is equivalent to the amount of data exchanged 
by WORLD TELECOM (today).
\end{itemize}
\clearpage
%
\section{TeV--scale Physics at LHC}
Exploiting the LHC physics potential means that we can 
answer or shed considerable light on fundamental open questions 
such as the mass problem or unification of fundamental interactions. 
First we discuss some issues related to parton luminosities,
i.e. the expected accuracy in SM cross--section measurements. This is 
important for
establishing signals for new physics which require a comparison 
of the measured SM cross--sections with those from beyond the 
SM processes.
The section on parton luminosities
is followed by the discussion of the Higgs sector and 
selected topics in sparticle searches
in order to  demonstrate the 
discovery potential of the proposed 
pp detectors\footnote{One usually assumes that  
``one'' LHC year with a peak luminosity of  
L=10$^{33}$cm$^{-2}$s$^{-1}$ and a running time of 
$10^{7}$s produces an integrated luminosity of 10 fb$^{-1}$. 
A more realistic estimate would use
an average run luminosity and includes losses
due to machine and detector 
efficiencies. It would thus be more conservative to assume that 
a running time of $10^{7}$s per year with the initial luminosity 
requires about 2--3 years to accumulate 10 fb$^{-1}$.}.
\subsection{Parton luminosities}

Accurate cross--section measurements for different 
SM processes are an important part of the LHC programme.
Previous studies have concluded that 
accuracies of  $\pm$5\% can be achieved.
These estimates are based essentially on 
the possibility to measure the proton--proton luminosity
and the subsequent uncertainties from the different 
parton distribution functions.
However, a different  proposed method might eventually 
lead to cross--section measurements which could approach $\pm$1\%
accuracies~\cite{lhclumi}. The basic idea of the proposed new
method is based on:
\begin{itemize}
\item
Experiments at the LHC will study the interactions 
between fundamental constituents of the proton, the quarks and 
gluons, at energies where these partons can be considered as 
quasi--free. Thus, the important quantity is the 
parton--parton luminosity 
at different values of $x_{\rm parton}$ 
and not the   
traditionally considered proton--proton luminosity.
\item
Assuming collisions of essentially free partons,
the production and decay of weak bosons, $u \bar{d} \rightarrow W^{+}
\rightarrow \ell^{+} \nu$,
$d \bar{u} \rightarrow W^{-}
\rightarrow \ell^{-} \bar{\nu}$
and $u \bar{u} (d \bar{d}) \rightarrow Z^{0} 
\rightarrow \ell^{+}\ell^{-}$ are, in lowest order, known
to at least a percent level.
Cross--section uncertainties from higher order QCD 
corrections are certainly larger, but are  
included in the measured weak boson event rates. 
Similar higher order QCD corrections to other 
$q\bar{q}$ scattering processes 
at different $Q^{2}$, like $q\bar{q} \rightarrow W^{+}W^{-}$,
can be expected. Thus, 
assuming that the $Q^{2}$--dependence can in principle
be calculated, very accurate theoretical predictions
for cross--section ratios like 
$\sigma(pp \rightarrow W^{+}W^{-})$/$\sigma(pp \rightarrow W^{\pm})$
should be possible. 
\item
It is a well known fact that 
the $W^{\pm}$ and $Z^{0}$ production rates at the LHC, 
including their leptonic branching ratios into electrons 
and muons, are huge and provide relatively clean  
and well measurable events with isolated leptons.
For instance, 
assuming L=10$^{33}$ cm$^{-2}$ s$^{-1}$, 
one expects per day about 10$^{6}$ W$\rightarrow l\nu$ events and about 
700 WW$\rightarrow l\nu l\nu$ events.
Using the well known $W^{\pm}$ and $Z^{0}$ masses,
possible $x$ values of quarks and antiquarks
are constrained by
m$_{W^{\pm}, Z^{0}}^{2} = s x_{q} x_{\bar{q}}$
with $s = 4 E_{beam}^{2}$. The product
$x_{q} x_{\bar{q}}$ at the LHC is therefore 
fixed to $\approx 3 \times 10^{-5}$. 
Thus, the rapidity distributions of the weak bosons 
are directly related to the fractional momenta $x$ 
of the quarks and antiquarks.
Consequently, the observable $\eta$--distributions of the 
charged leptons from the decays of $W^{\pm}$ and $Z^{0}$ bosons
are also related to the $x$--distributions of 
quarks and antiquarks.
The shape and rate of the lepton $\eta$--distributions
provide therefore the key to precisely constrain the 
quark and antiquark structure functions 
and their corresponding luminosities.   
\end{itemize}

Using a 
PYTHIA simulation  it could be shown that the rapidity 
distributions of $W^{+}$, $W^{-}$ and $Z^{0}$
events, identified through their clean leptonic decays,
determine directly and very accurately 
the $x$--distribution of quarks and antiquarks 
and their corresponding luminosity.
The sensitivity of these lepton distributions 
to recent parton distribution functions is demonstrated in
figure 15 for W--decays. Furthermore, it was shown that
cross--sections of other $q\bar{q}$ related processes are
strongly correlated with the 
single $W^{\pm}$ and $Z^{0}$ production. 
Ignoring remaining theoretical uncertainties from 
missing higher order calculations, one finds that 
ratios like $\sigma(q\bar{q} \rightarrow W^{+}W^{-})
/\sigma(q\bar{q} \rightarrow W^{\pm})$ show stability 
within better than 1\%. 
Thus, the 
shape and height of $W^{\pm}$ and $Z^{0}$ rapidity distributions
provide a precise LHC luminosity monitor 
for $q, \bar{q}$ parton $x$--distributions 
at a $Q^{2}$ = m$_{W, Z}^{2}$.
A similar analysis using qg$\rightarrow \gamma$ + jet and  qg$\rightarrow$ 
Z + jet 
has been performed, and shows that the $x$--distribution of
gluons can be extracted with similar precision~\cite{gluon}.
\begin{figure}[htb]
\begin{center}
\epsfig{file=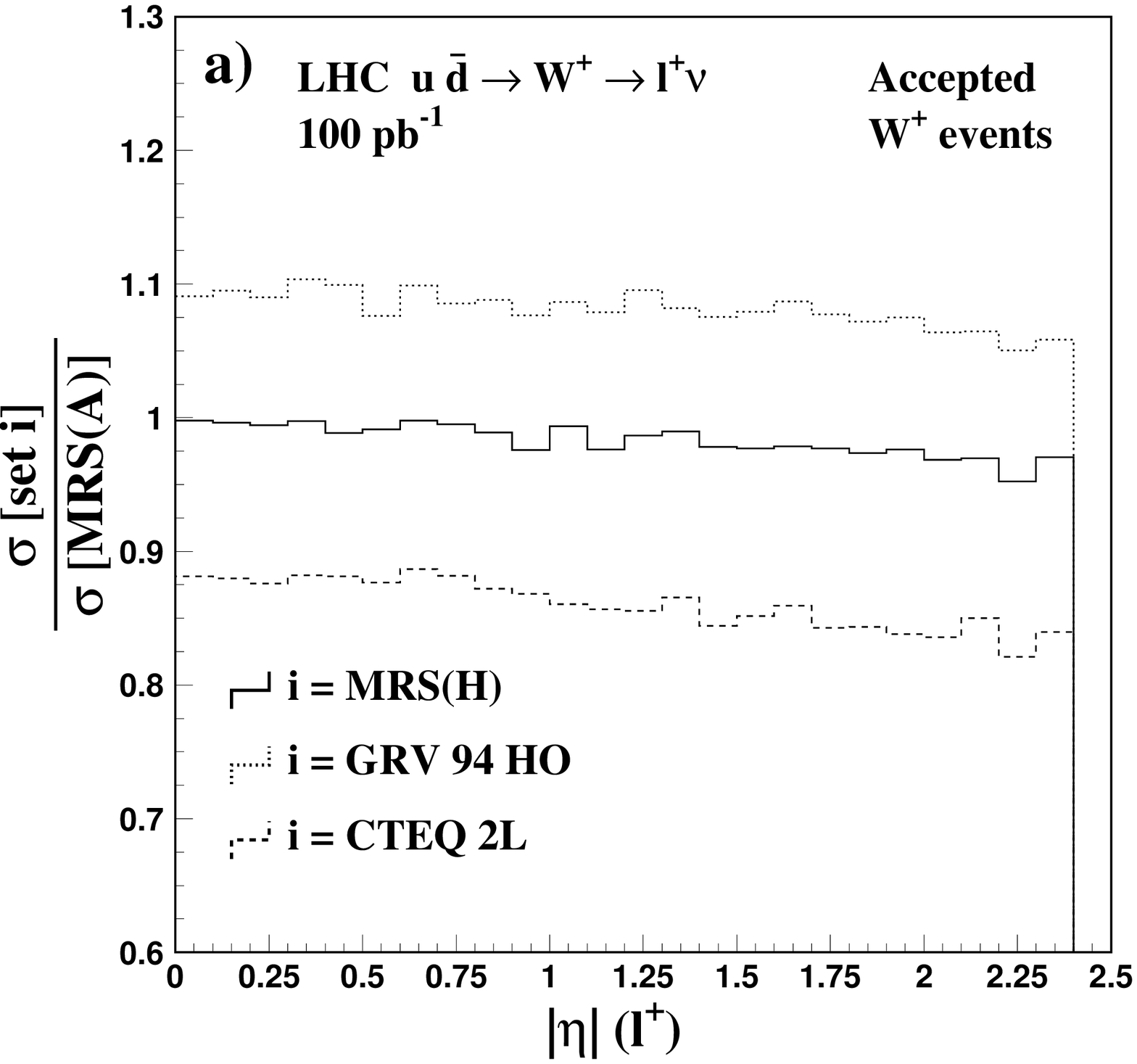,
height=6.cm,width=7.cm}
\epsfig{file=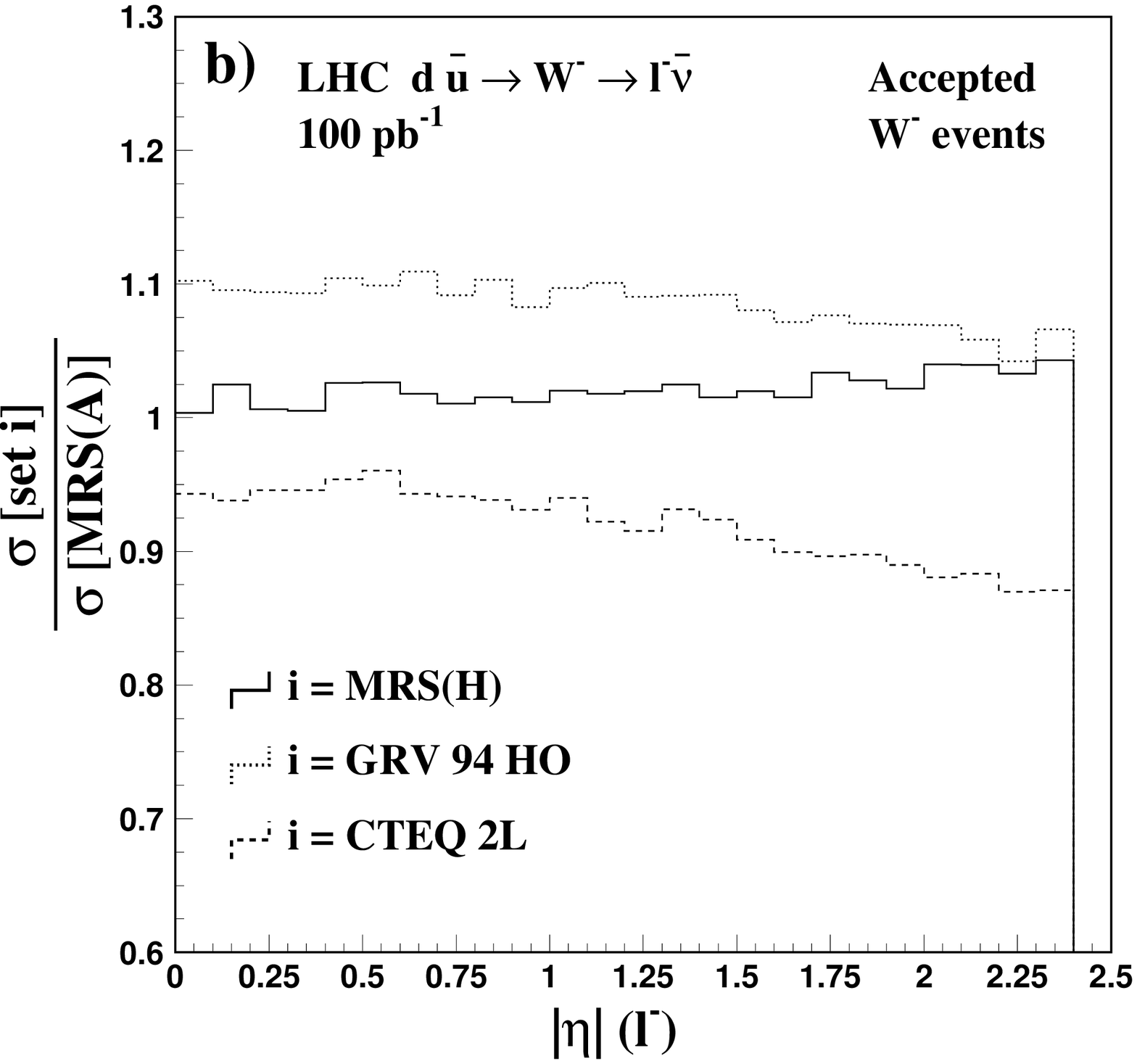,
height=6. cm,width=7.cm}
\end{center}
\caption[fig15]
{Rapidity dependence of the $\ell{^\pm}$ 
cross--section predictions for $W^{\pm} \rightarrow \ell^{\pm} \nu$
with different sets of structure functions relative to the one 
obtained from the MRS(A) parametrisation; a) for $\ell^{+}$,
b) for $\ell^{-}$~\cite{lhclumi}.}
\end{figure}
\clearpage 

\subsection {SM Higgs search }

Figure 16 shows the next--to--leading order   
Higgs cross--sections~\cite{zoltan} at the LHC for 
various production processes as a function of the Higgs mass.
By far the largest contribution comes from the 
gluon--gluon fusion process~\cite{georgi}. Depending on the Higgs mass, its
detection involves several different signatures. 
The Higgs search is therefore an excellent reference physics process 
to evaluate the overall detector performance. 
In particular, the search for the intermediate Higgs 
(m$_{Z} \leq$ m$_{H} \leq $2m$_{Z}$) is known to pose demanding 
requirements on the detectors. The natural width of the Higgs in 
this mass range is very small. The measured width of the signal 
will therefore be dominated entirely by the instrumental mass resolution. 
\begin{figure}[htb]
\begin{center}
\includegraphics*[scale=0.5,angle=-90,bb=50 0 480 765 ]
{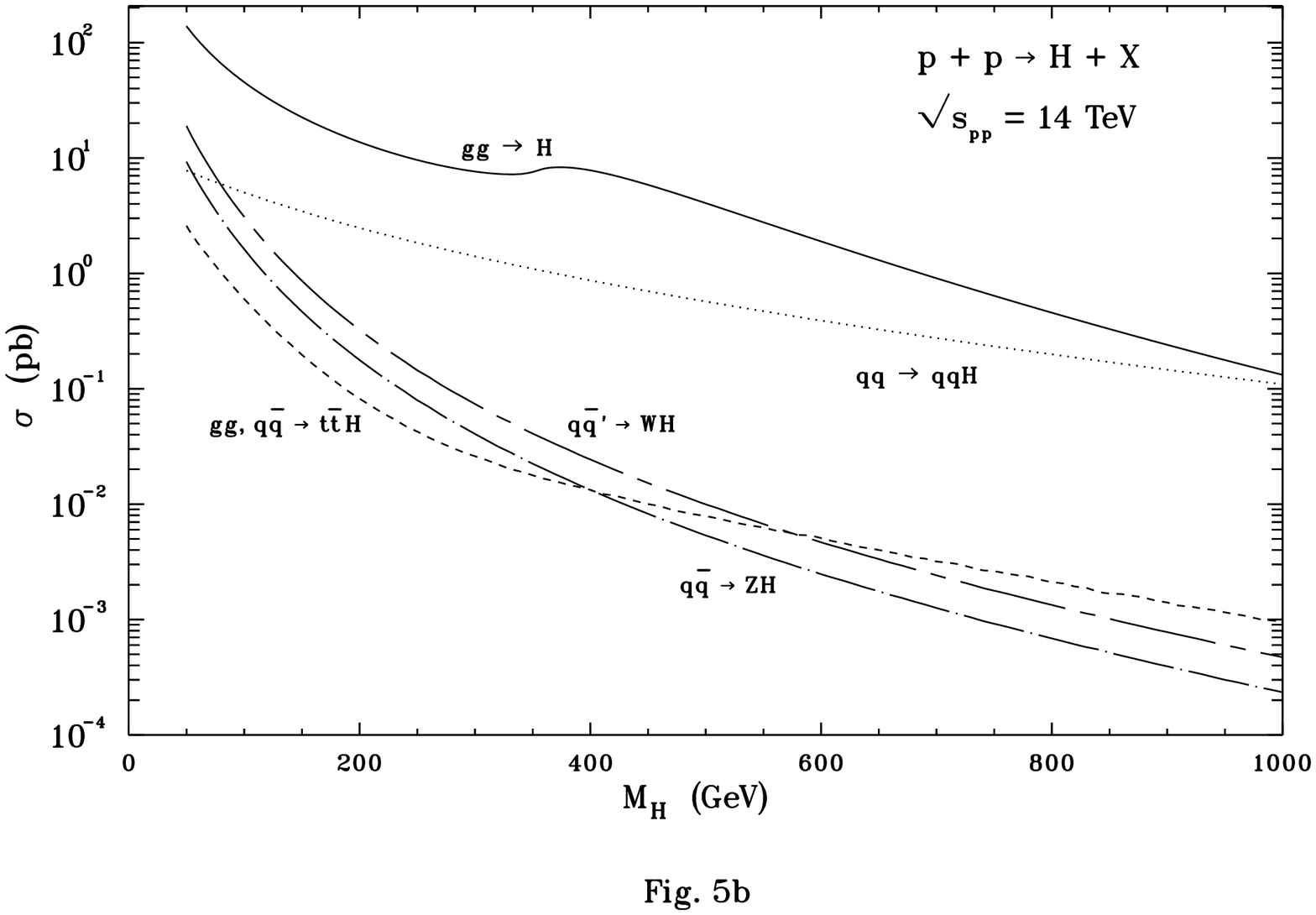}
\caption[fig16]
{Next--to--leading order cross--section calculations  for the SM 
Higgs~\cite{zoltan}.} 
\end{center}
\end{figure}
\begin{figure}[htb]
\begin{center}
\epsfig{file=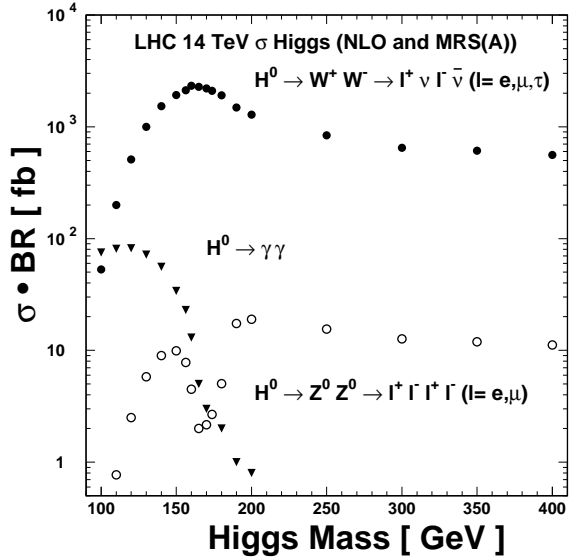,width=8.cm}
\end{center}
\caption[fig17]
{Expected $\sigma \times BR$ for different detectable 
SM Higgs decay modes~\cite{hdecays}.} 
\end{figure}
Figure 17 shows the $\sigma \times BR$~\cite{hdecays} for 
the most promising Higgs search channels: 
$H \rightarrow \gamma \gamma$,
$H \rightarrow Z Z^{(*)} \rightarrow 4 \ell^{\pm}$,
and $H \rightarrow W W^{(*)} 
\rightarrow \ell^{+} \nu \ell^{-} \bar{\nu}$.

Figure 18 summarises the expected observability of the 
SM Higgs in ATLAS\footnote{The expected Higgs signal significance
from ATLAS does not yet include the 
channel $H \rightarrow W W^{(*)} 
\rightarrow \ell^{+} \nu \ell^{-} \bar{\nu}$.}~\cite{atlasperf1} 
and CMS~\cite{cmshiggs1}, assuming 100 fb$^{-1}$.

\begin{figure}[htb]
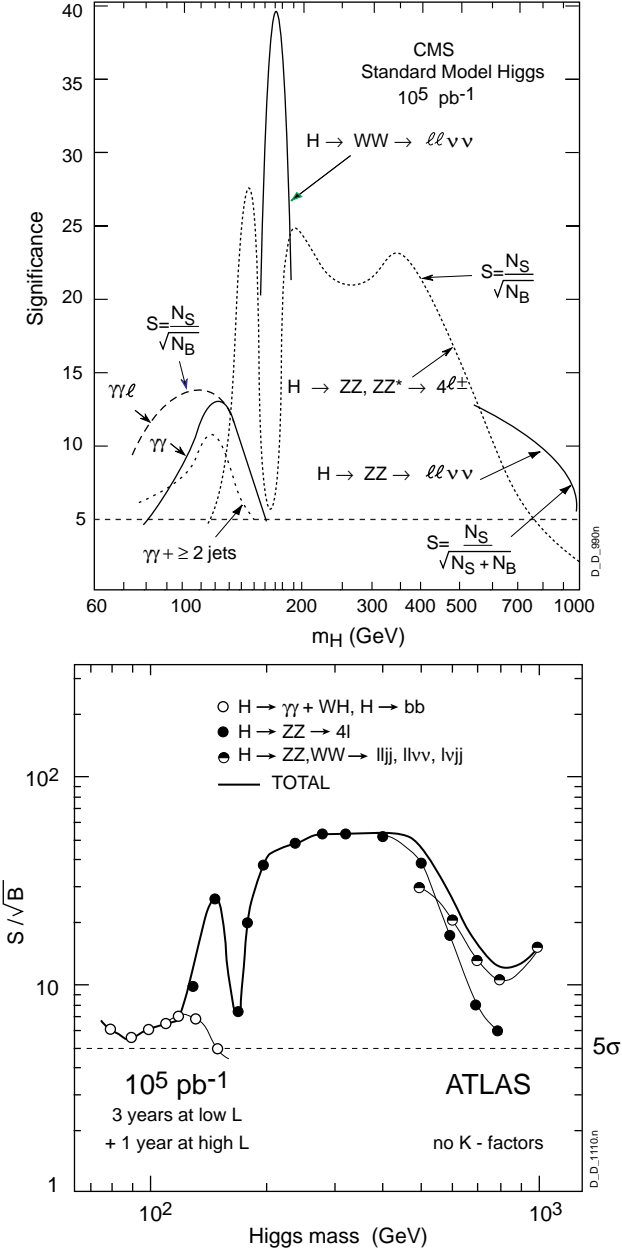

\begin{center}
\includegraphics*[scale=0.5,bb=79 92 516 590]{D_Denegri_0990n.ill}
\includegraphics*[scale=0.5,bb=68 167 535 617]{D_Denegri_1110n.ill}
\end{center}
\caption[fig18]{Expected signal significances for 
the SM Higgs search in ATLAS~\cite{atlasperf1} and CMS~\cite{cmshiggs1},
assuming 100 fb$^{-1}$.}
\end{figure}

The most promising signature for a SM Higgs 
with masses between the expected LEP200 limit 
and 130 GeV is the decay $H \rightarrow \gamma \gamma$ 
with a branching ratio of only $\approx 2 \times 10^{-3}$.
As can be seen from figure 19,  
this signal has to be detected above a 
large background from continuum $\gamma \gamma$ events. 
The detection of such a signal requires an
excellent $\gamma \gamma$ mass resolution of 
$\leq$ 1\% (i.e. $\leq$ 1 GeV for m$_{H}$ = 100 GeV and 
a very good  $\pi^{0}$ rejection capability.
More details about this channel can be found 
in section 6.2.1.

\begin{figure}[htb]
\begin{center}
\mbox{
\epsfig{file=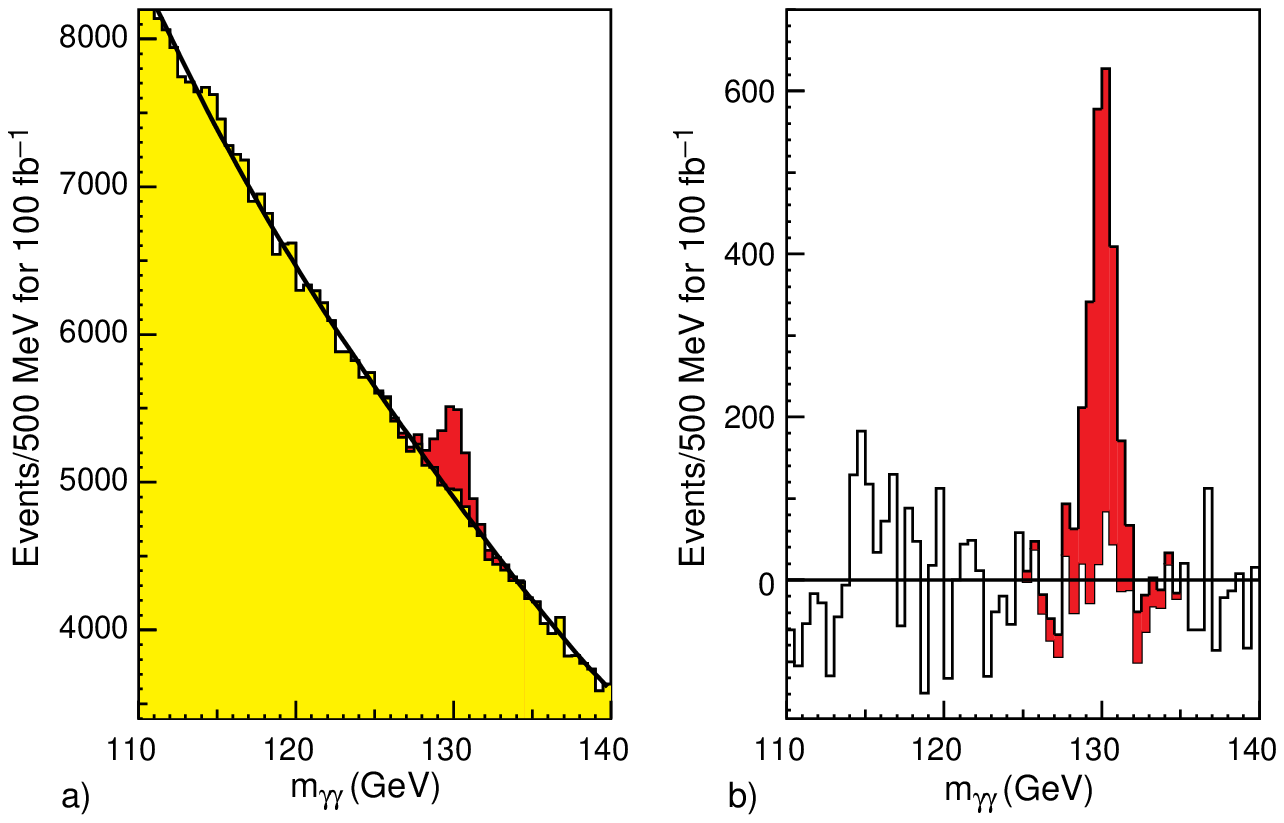,
height=6 cm,width=12cm}
}
\end{center}
\caption[fig19]
{CMS  simulation for 
$H \rightarrow \gamma \gamma$ (m$_{H}$ = 130 GeV) before and after background 
subtraction~\cite{ecaltdr}.}  
\end{figure}
%
For Higgs masses between 130 GeV and 200 GeV
the sensitivity of the $4 \ell^{\pm}$ signature suffers from very low 
branching ratios as illustrated in figures 17 and much  
smaller signals, like the ones shown in figure 20, are expected.
Consequently, a 5 standard deviation signal  
requires integrated luminosities of at least 30--100 fb$^{-1}$.
A recent study has demonstrated that this Higgs mass region 
can also be covered by the
$H \rightarrow W W(^{*}) \rightarrow
\ell^{+} \nu \ell^{-} \bar{\nu}$ decay~\cite{hww1}. 
The performed analysis, described in section 6.2.2,
shows that this channel should allow to discover 
a SM Higgs with 5 standard deviation for a Higgs mass 
between 140--200 GeV and integrated luminosities below 5 fb$^{-1}$. 
\begin{figure}[htb]
\begin{center}
\epsfig{file=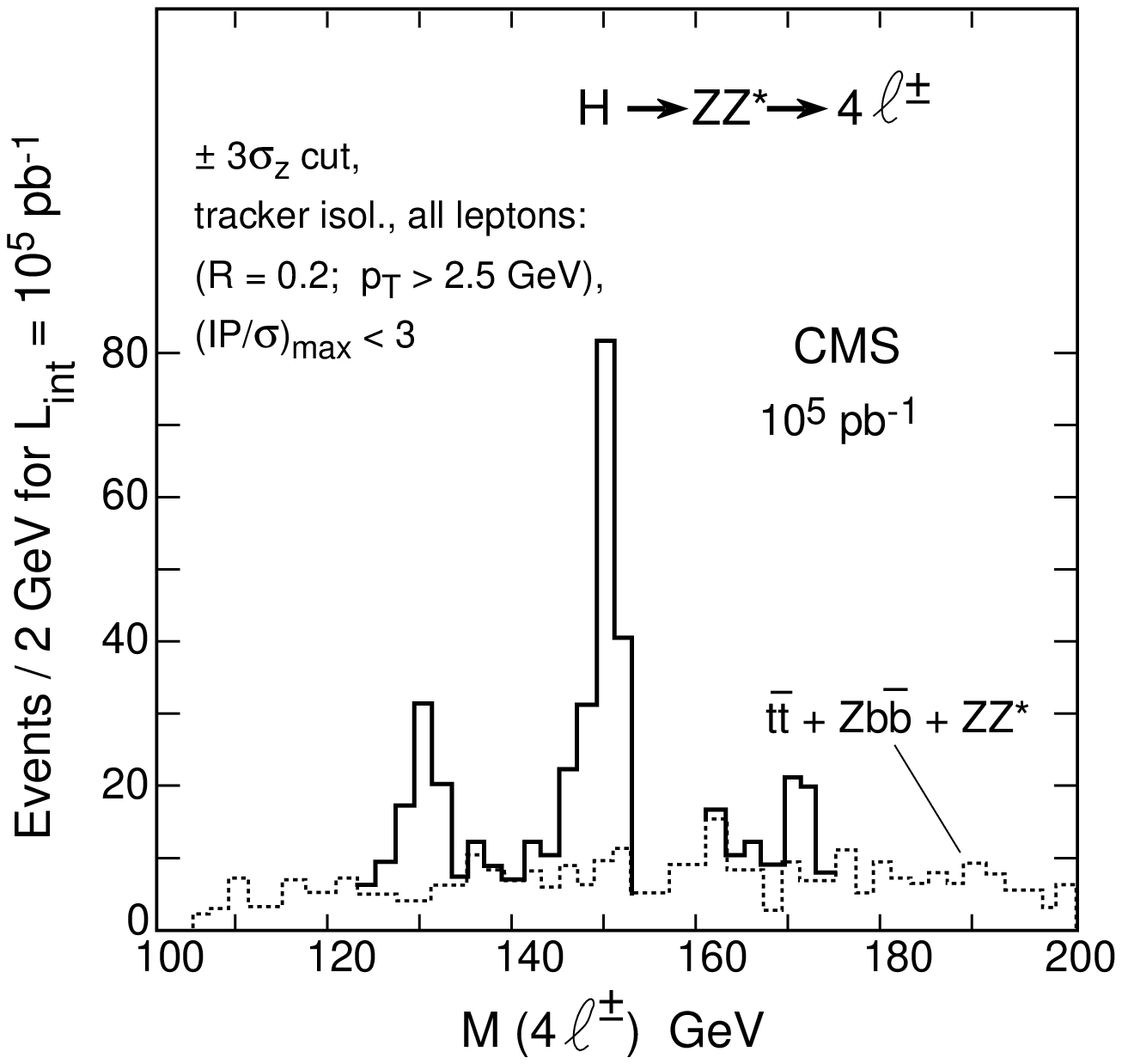,  
bbllx=70pt,bblly=240pt,bburx=500pt,bbury=630pt,
height=8.cm,width=11.cm}
\end{center}
\caption[fig20]
{CMS simulation for $H \rightarrow Z Z^{*} \rightarrow
\ell^{+}\ell^{-} \ell^{+}\ell^{-}$ and m$_{H}=130, 
150$ and $170$ GeV.
}
\end{figure}

For 2$\times$m$_{Z} \leq$ m$_{H} \leq$ 500 GeV the 
decay $H \rightarrow Z Z \rightarrow 4\ell^{\pm}$ 
provides the experimentally easiest discovery signature as
the events should contain four isolated high p$_{T}$ leptons. 
A CMS simulation of a Higgs search using the 
4 lepton invariant mass distribution is shown in figure 21, 
obvious Higgs mass peaks are visible. 
Furthermore, a Z mass constraint can be used for both lepton pairs
to suppress other backgrounds.
Estimates from ATLAS and 
CMS indicate that an integrated luminosity of about
10 fb$^{-1}$ is required to discover a SM Higgs in this mass range
with at least 5 standard deviation~\cite{atlascms}. 
For example, an ATLAS study~\cite{atlasnote1} shows that
a Higgs (m$_{H}=300$ GeV and 
$H \rightarrow ZZ \rightarrow 4 \ell^{\pm}$) 
should be seen with 
35 signal events above a continuum 
background of $\approx 13 \pm 4$ events, assuming 10 fb$^{-1}$.
This study indicates also that the signal--to--background 
rate can be significantly improved by requiring
that one reconstructed $Z$ has a p$_{T} \geq$ m$_{H} / 2$.
Using this cut results in 13 signal events (m$_{H}$ = 300 GeV) and 
a background of 0.6 events  (10 fb$^{-1}$).
\begin{figure}[htb]
\begin{center}
\epsfig{file=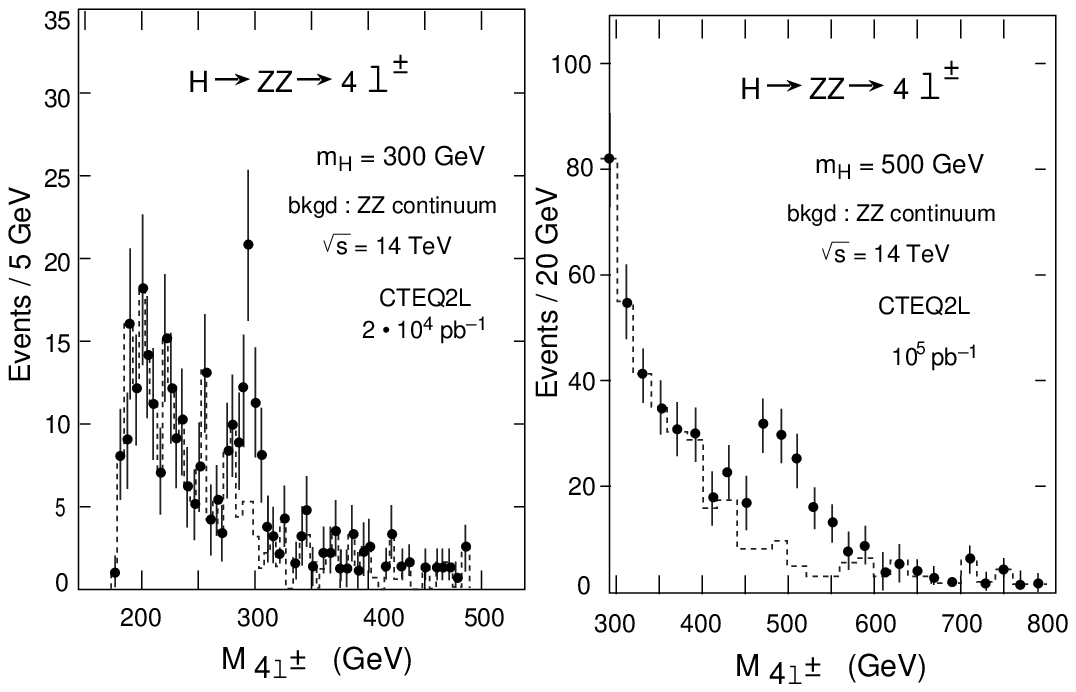,  
bbllx=150pt,bblly=380pt,bburx=460pt,bbury=580pt,width=\textwidth,clip=}
\end{center}
\caption[fig21]
{CMS simulation results for $H \rightarrow Z Z \rightarrow
\ell^{+}\ell^{-} \ell^{+}\ell^{-}$ and m$_{H} = 300$ GeV
and m$_{H} = 500$ GeV.
}
\end{figure}

For Higgs masses above $\approx$ 400 GeV additional 
signatures involving hadronic $W$ and $Z$ decays 
as well as invisible $Z$ decays like
$H \rightarrow Z Z \rightarrow \ell^{+}\ell^{-} \nu \bar{\nu}$ 
have been investigated (see figure 22). The advantages of much larger 
branching ratios are however spoilt
by larger backgrounds from  $t\bar{t}$, $W + X$ and $Z + X$. 
These high mass Higgs signatures involve missing transverse energy 
and jet--jet masses and require thus hermetic detectors with
good jet--energy reconstruction.

\begin{figure}[htb]
\begin{center}
\epsfig{file=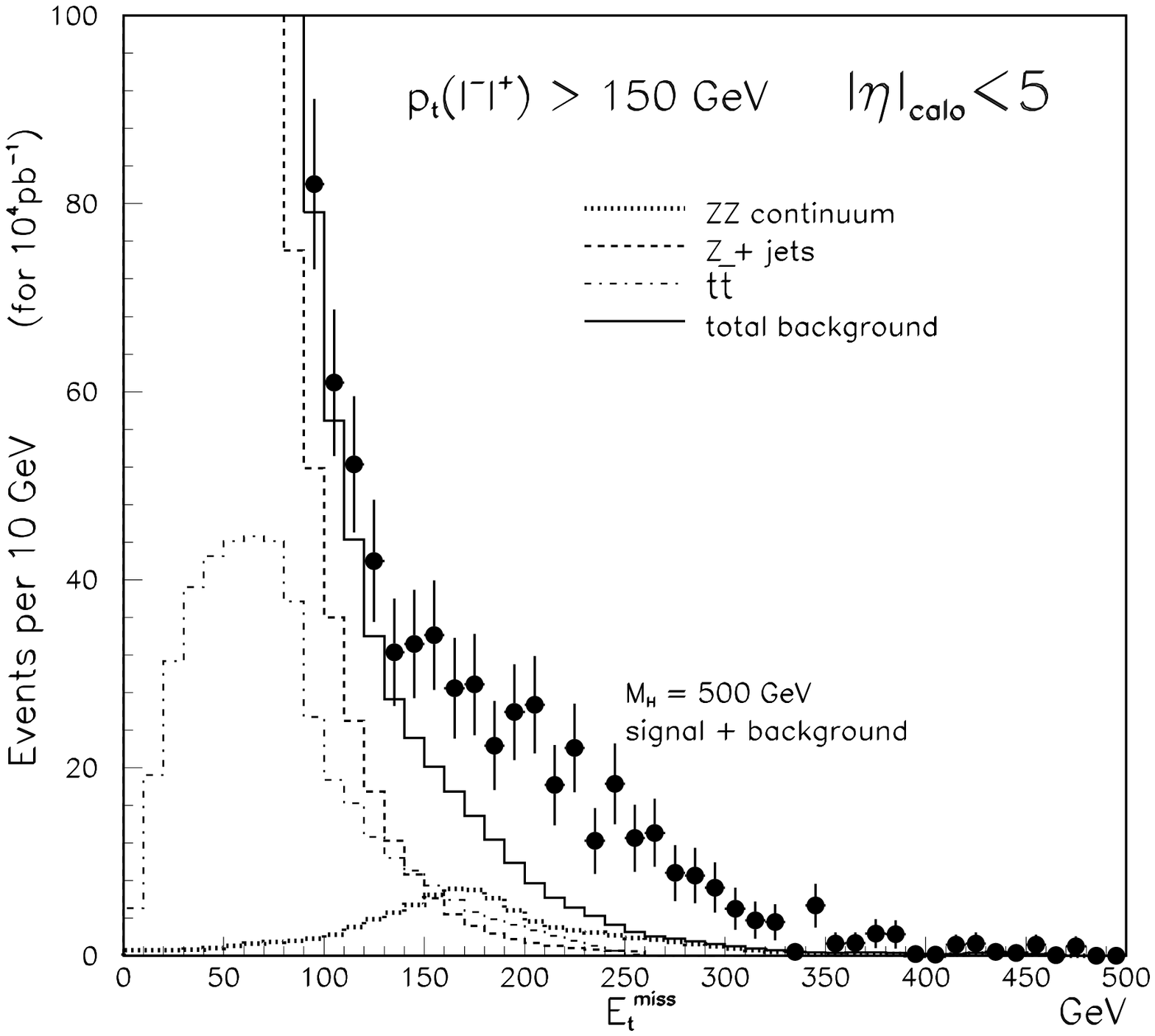,width=6.0cm}
\epsfig{file=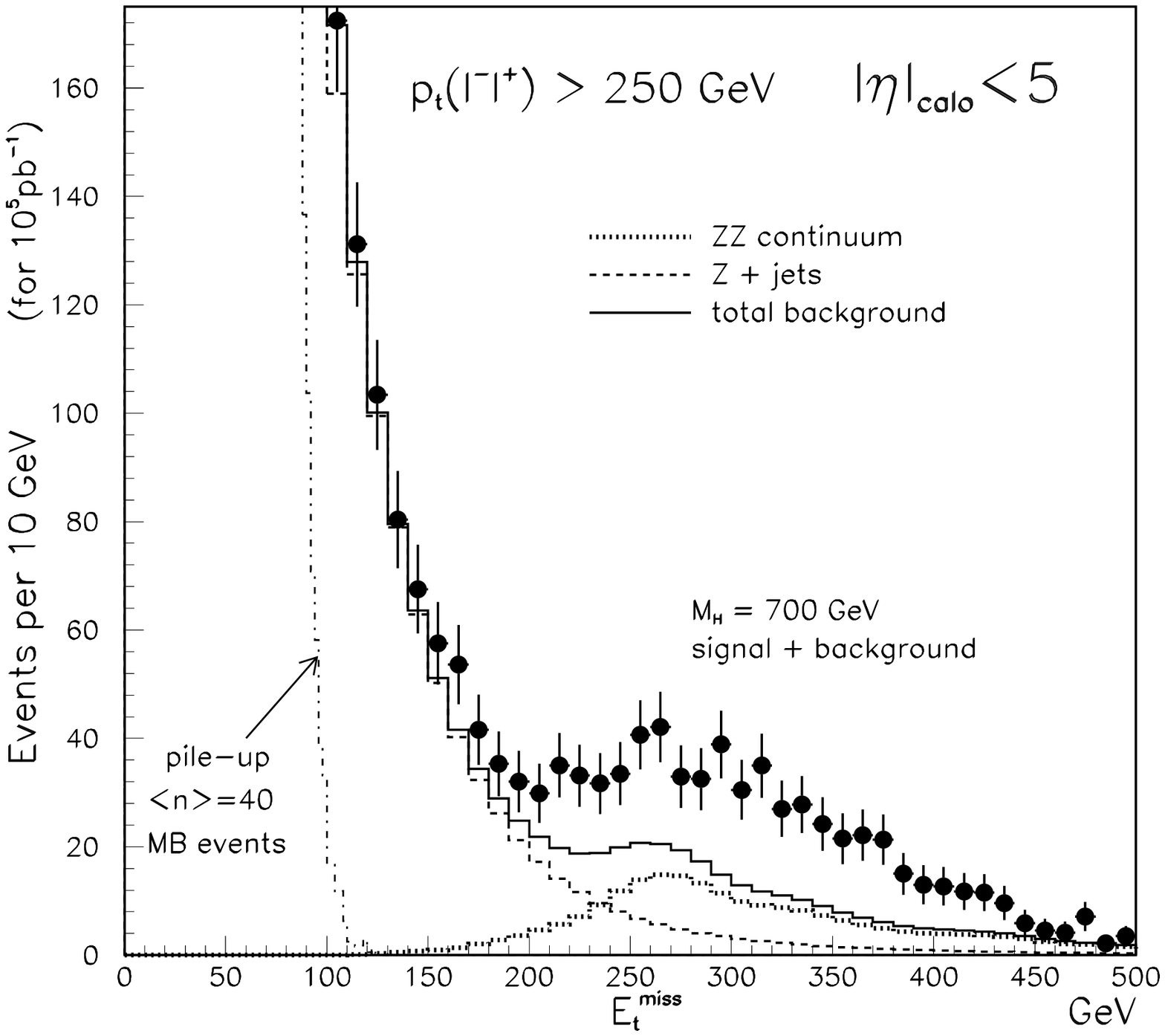,width=6.0cm}
\end{center}
\caption[fig22]
{ATLAS simulation results for $H \rightarrow Z Z \rightarrow
\ell^{+}\ell^{-} \nu \bar{\nu}$.}
\end{figure}

\clearpage
\subsubsection{The SM $H \rightarrow \gamma \gamma $ channel}

The $\gamma \gamma$ mass resolution 
depends upon the energy resolution and the resolution on the 
measured angle between the two photons. As regards the angle 
between the photons, the issue is the possible uncertainty 
on the knowledge of the position of the production vertex. 
Although very localised in the transverse plane, 
the interaction vertices have a r.m.s. spread of about 53 mm 
along the beam axis. 
If no other knowledge were available such a spread would contribute 
about 1.5 GeV to the mass resolution. Detailed studies suggest 
that the correct vertex can be located using charged tracks,
even at the highest luminosities, 
where there are on average nearly 20 inelastic interactions 
per bunch--crossing. This method of using tracks for the vertex 
localisation is based on the expectation that the Higgs 
production events are harder than minimum--bias events and that they 
contain more high--p$_{T}$ tracks. Using this fact it is possible to 
devise an algorithm to select the vertex of the Higgs event 
from the background of other primary vertices in the same bunch--crossing. 
This method is used by CMS~\cite{ecaltdr}.

ATLAS can use in addition the 1$^{st}$ and 2$^{nd}$ sampling of the 
electromagnetic calorimeter to measure the photon direction. 
In this case a contribution to the $\gamma \gamma$--mass resolution 
of about 530 MeV is expected~\cite{atlasecal}.
\begin{figure}[htb]
\begin{center}
\epsfig{file=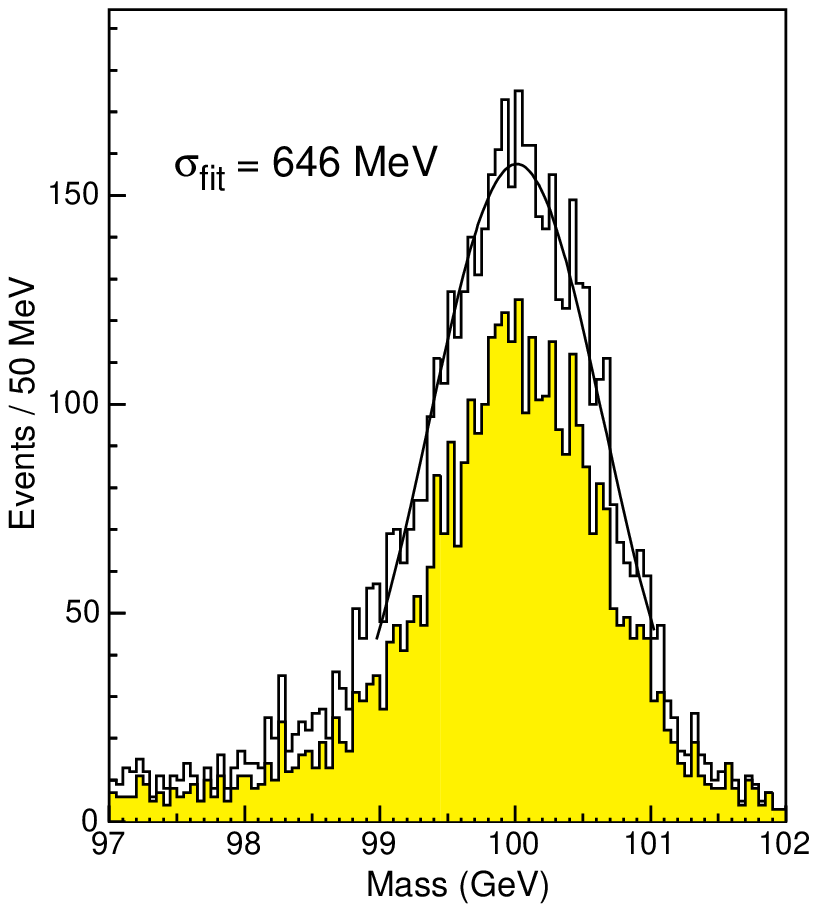,height=7.0cm}
\epsfig{file=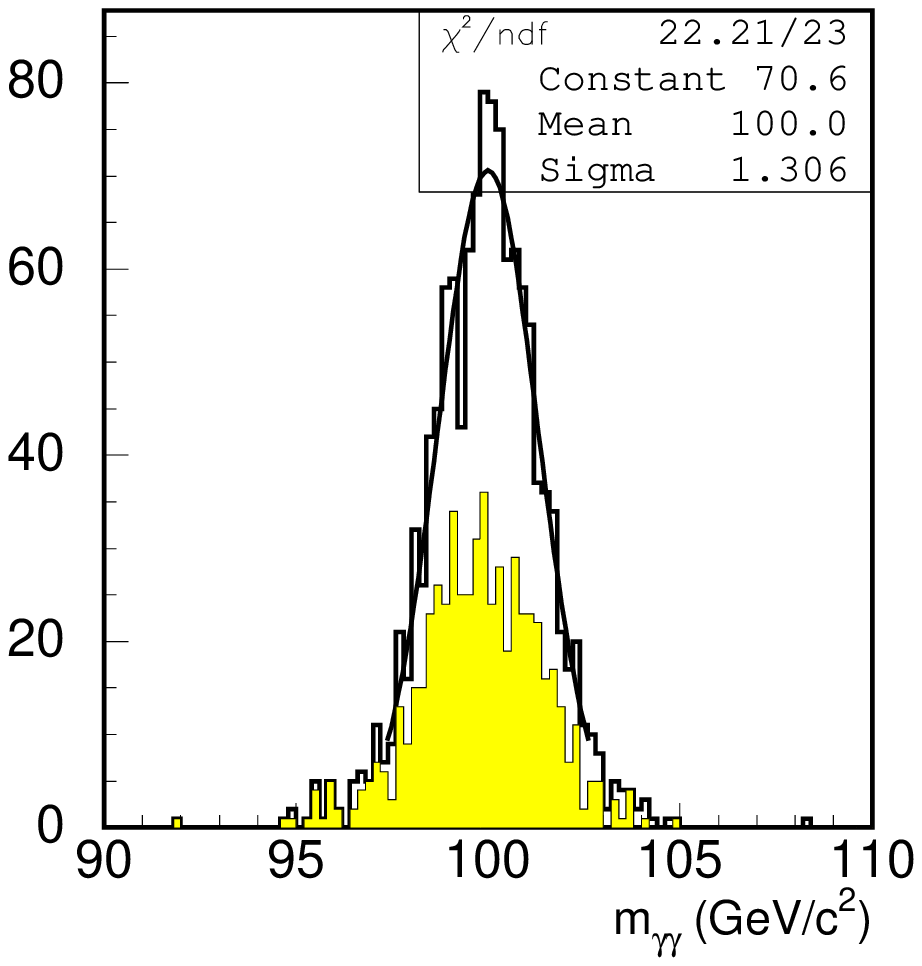,height=7.0cm}
\end{center}
\caption[fig23]
{Higgs mass resolution including converted photons: 
in the left plot (CMS)~\cite{ecaltdr}
the unconverted photons are shown as shaded area and in 
the right plot (ATLAS)~\cite{atlasperf}
the shaded area corresponds to the converted photons}
\end{figure}

Due to the material in front of the electromagnetic calorimeter 
(beam--pipe, inner tracking detector with support structures) photons 
will convert. In both experiments about 50\% of 
the $H \rightarrow \gamma \gamma$ 
events have one or both of the photons converted.
Detailed simulation studies 
have shown that  a large fraction of these converted photons can be 
recovered with only a small degradation in resolution. 
Figure 23 shows the Higgs mass resolution taking the 
converted photons into account. 
This figure shows further that CMS expects about a factor 2 
better mass resolution than ATLAS, which demonstrates 
the potential superior performance of a crystal calorimeter.

The dominant jet--background to the  $H \rightarrow \gamma \gamma$ signal 
comes from jet--$\gamma$ events, where the jet fragments to a 
leading $\pi ^{0}$,
carrying a large fraction of the jet transverse momentum. 
Isolation criteria using calorimeter and/or charged tracks are very 
powerful tools to reduce this potentially large background. 
In addition isolated  $\pi ^{0}$s can be rejected by detecting 
the presence of two close--by electromagnetic showers rather than one.
This can be achieved 
 using the lateral shower shape of the electromagnetic cluster. 
The resulting rejection factor depends strongly on the $\pi ^{0}$ 
transverse momentum, e.g. rejection factors larger than 3 for 
p$_{T} < $ 40 GeV can be achieved with a small $\gamma$--efficiency loss.
Figure 24 illustrates that isolation cuts together with 
a $\pi^{0}$--rejection algorithm reduce 
the $\gamma$--jet background well below the 
intrinsic $\gamma\gamma$ background. 
\begin{figure}[htb]
\begin{center}
\epsfig{file=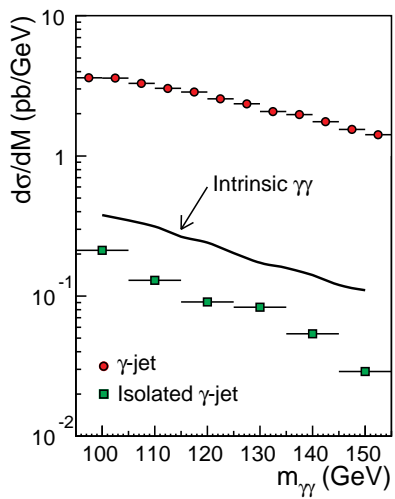,height=7cm}
\end{center}
\caption[fig24]
{$\gamma$--jet background cross--section as a function of mass before 
and after isolation. The line shows the level of the irreducible 
di--photon background expected in CMS~\cite{ecaltdr}.}
\end{figure}

The signal significance (N$_{S}$/$\sqrt{N_{B}}$) for a SM Higgs decaying 
to two photons has been evaluated using events within a  $\pm$ 1.4 $\sigma$ 
mass window. Figure 25 shows the expected 
signal significance from CMS~\cite{ecaltdr}, 
as a function of the Higgs mass, for 30 fb$^{-1}$ and 100 fb$^{-1}$.  
This figure demonstrates further that 
a luminosity of 30 fb$^{-1}$ should enable CMS to
detect the Higgs in the mass 
range between 100-150 GeV with more than five standard deviation
in the decay $H \rightarrow \gamma \gamma$.

\begin{figure}[htb]
\begin{center}
\epsfig{file=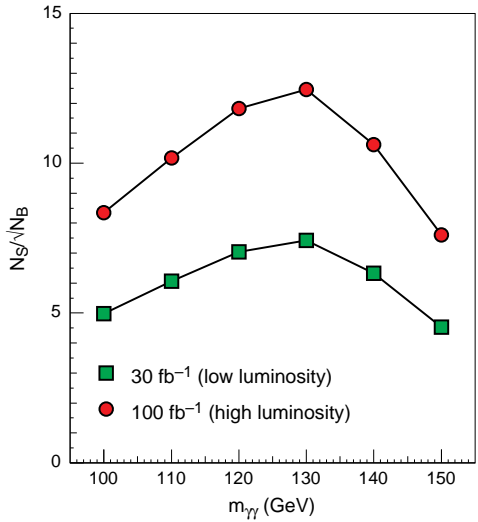,height=7.0cm}
\end{center}
\caption[fig25]
{Signal significance as a function of m${_H}$, for 
H$\rightarrow \gamma \gamma$ seen after 30 fb$^{-1}$ and 
100 fb$^{-1}$ collected in CMS at low and high luminosity 
respectively~\cite{ecaltdr}.}
\end{figure}
%
%
\subsubsection{The SM $H \rightarrow W^{+}W^{-} \rightarrow 
\ell^{+} \nu \ell^{-} \bar{\nu} $ channel}

A recent simulation has demonstrated that the  
$H \rightarrow W^{+}W^{-} \rightarrow 
\ell^{+} \nu \ell^{-} \bar{\nu}$  channel can be used to observe a 
statistically significant signal in the Higgs mass range of 130--200 GeV.
This analysis~\cite{hww1} exploits two important differences 
between a Higgs signal
and the non--resonant background from $pp \rightarrow W^{+}W^{-} X$.
\begin{figure}[htb]
\begin{center}\mbox{
\epsfig{file=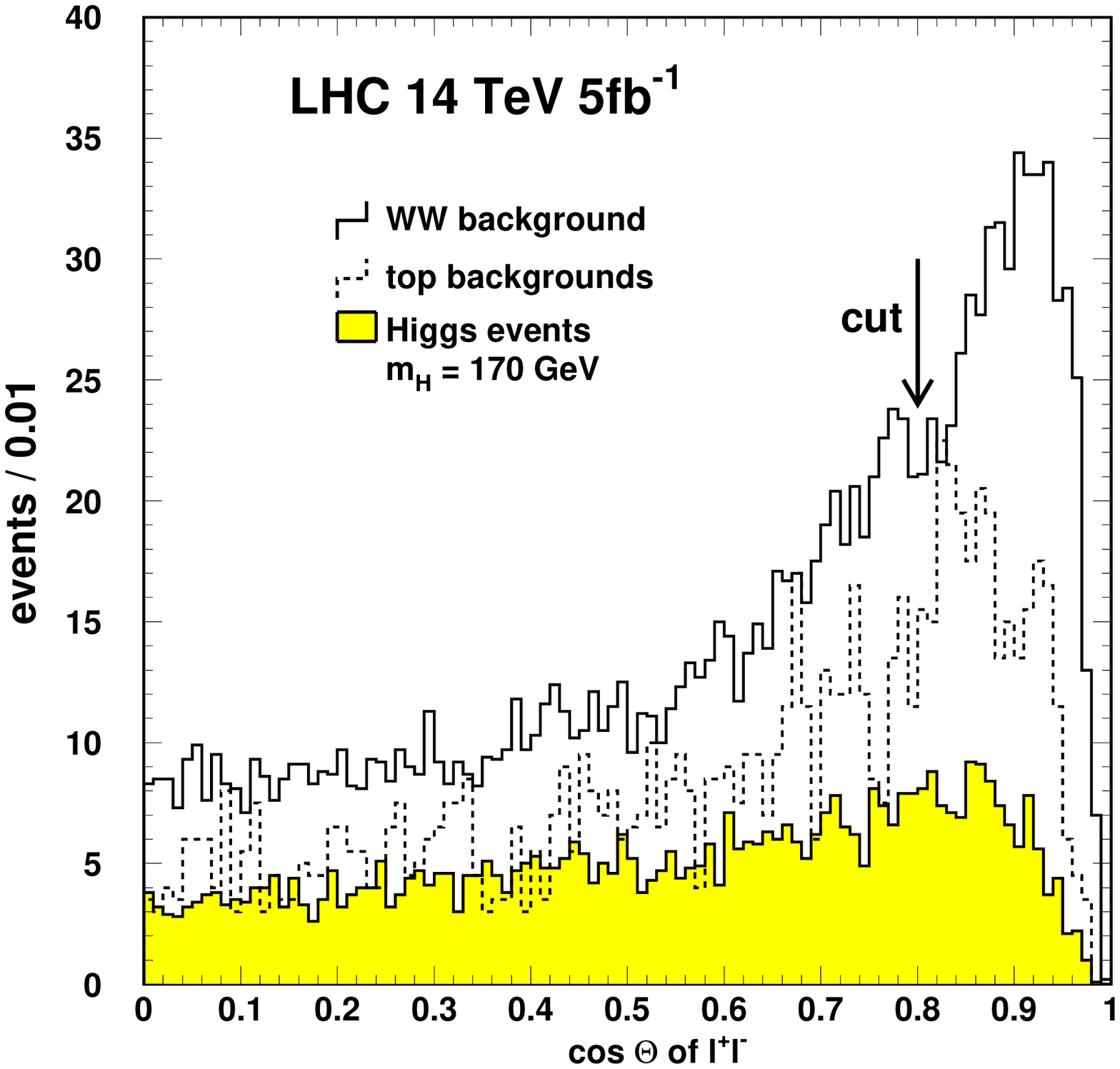,
height=7 cm,width=7cm}
\epsfig{file=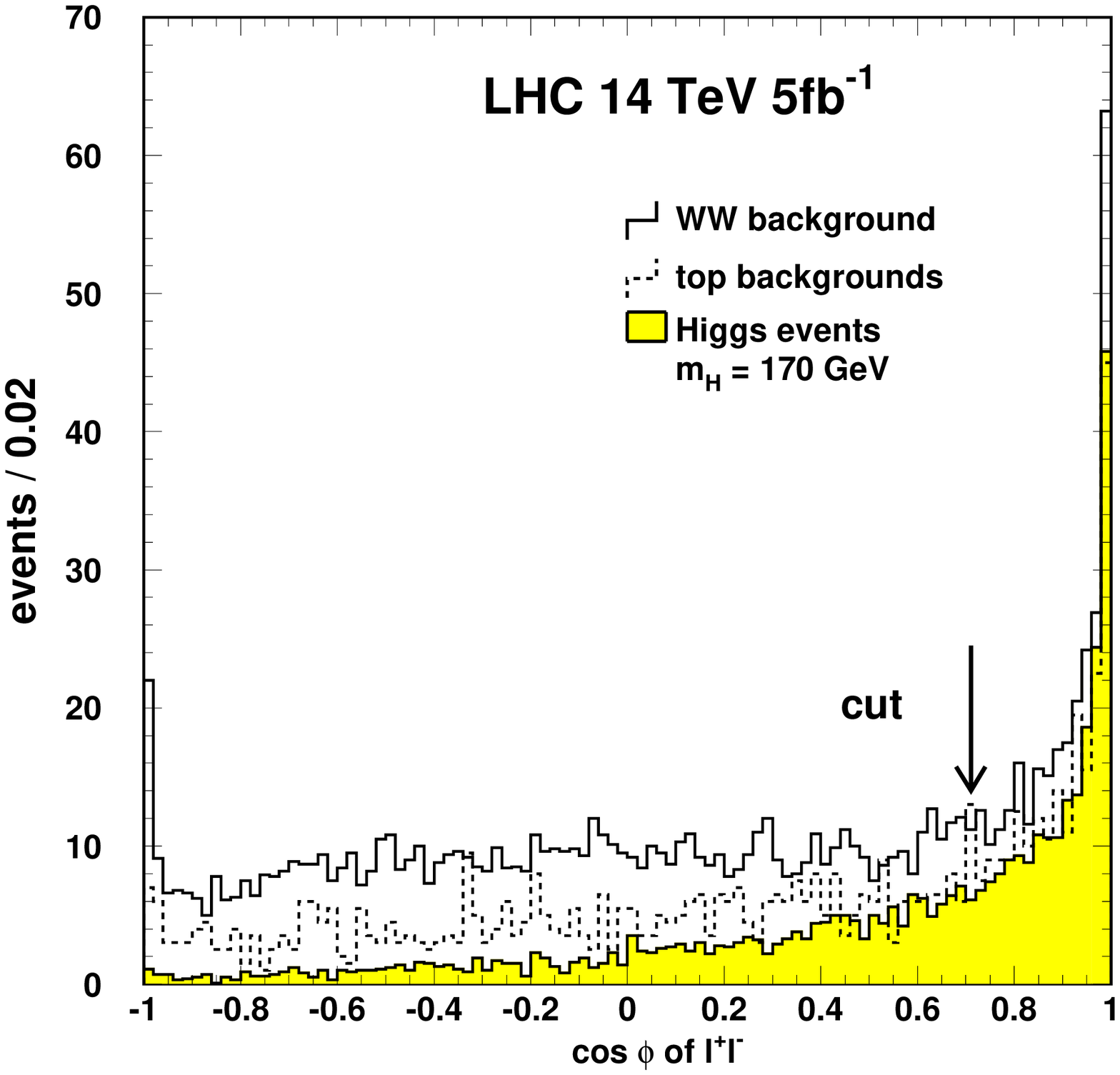,
height=7 cm,width=7cm}}
\end{center}
\caption[fig26]{$| \cos \theta |$ 
distribution of the dilepton  
system with respect to the beam direction and $ \cos \phi $ 
distribution of the dilepton system in the plane transverse
to the beam direction for Higgs signal 
and background events~\cite{hww1}. 
}
\end{figure}
\begin{figure}[htb]
\begin{center}\mbox{
\epsfig{file=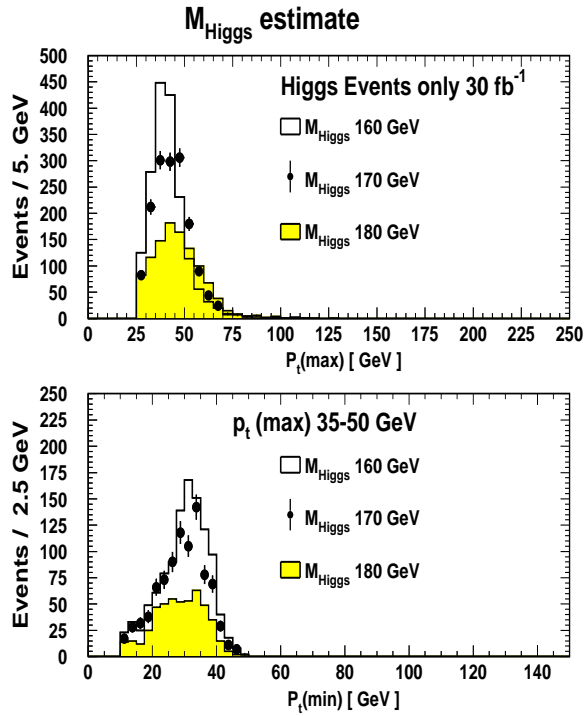,
width=8.cm,height=10.cm}}
\end{center}
\caption[fig27]
{Expected lepton p$_{T}$ spectra for 
$H\rightarrow W^{+}W^{-}\rightarrow \ell^{+} \nu \ell^{-} \bar{\nu}$ 
and three different Higgs masses~\cite{hww1}.}
\end{figure}
As shown in figure 26, the 
signal events from gluon--gluon scattering are  
more central than the $W^{+}W^{-}$ 
background from $q\bar{q}$ scattering.
This difference is exploited by the requirement that 
the polar angle $\theta$ 
of the reconstructed dilepton momentum vector,
with respect to the beam direction, satisfies $|\cos \theta| < 0.8$.
As a result, both leptons are found essentially within the 
barrel region ( $|\eta| < 1.5$) of the experiments.
The $ \cos \phi $ distribution in 
figure 26 shows the effect of $W^{+}W^{-}$ spin correlations
and the V--A structure of the $W$ decays which results in a 
distinctive signature for $W^{+}W^{-}$ pairs produced in
Higgs decays. For a Higgs mass close to $2 \times$m$_{W}$, the 
W$^{\pm}$ boost is small and the opening 
angle between the two charged leptons in the plane 
transverse to the beam direction is very small.

Finally, the lepton p$_{T}$ spectra,
which are sensitive to the Higgs mass as shown
in figure 27, can further be used to improve the signal 
to background ratio and to 
determine the Higgs mass with an accuracy of 
$\delta$m$_{H} \approx \pm$ 5 GeV, assuming 5 fb$^{-1}$.
%
%

\clearpage
\subsection { SUSY Searches}

The attractive features of the MSSM are very well described 
in a Physics Report  by H. P. Nilles~\cite{nilles} in 1984. 
We repeat here some of his 
arguments given in the introduction of the report: \newline

{\it ``Since its discovery some ten years ago, Supersymmetry has 
fascinated many physicists. This has happened despite the absence of 
even the slightest phenomenological indication that it might be relevant 
for nature. .... Let us suppose that the Standard Model is valid up 
to a grand unification scale or even the Planck scale of $10^{19}$ GeV.
The weak interaction scale of 100 GeV is very tiny compared to these 
two scales. If these scales were input parameters of the theory 
the (mass)$^2$ of the scalar particles in the Higgs sector have to 
be chosen with an accuracy of $10^{-34}$ compared to the Planck Mass. 
Theories where such adjustments of incredible accuracy have to 
be made are sometimes called unnatural.... 
Supersymmetry might render the Standard Model natural... 
To render the Standard Model supersymmetric a price has to be paid. For 
every boson (fermion) in the Standard Model, a supersymmetric partner 
fermion (boson) has to be introduced and to construct phenomenological
acceptable models an additional Higgs supermultiplet is needed.''}

SUSY signatures are excellent benchmark
processes to evaluate the
physics performance of LHC 
detectors and they thus have influenced the detector optimisation.
In order to cover the  largest possible parameter space in the Higgs sector
the searches are  more challenging compared to the  SM Higgs because: (i)
one low mass Higgs (h) {\em must} exist, (ii) there are 5 Higgs bosons:
h, H$^{0}$, A$^{0}$ , H$^{\pm}$ and (iii)
the expected ($\sigma \cdot$ BR)$_{SUSY}$ for the 
$\gamma \gamma$- and $4l$--channel are smaller than for the 
SM Higgs.
For the simulation results discussed below, all sparticle masses 
are assumed to be heavy enough such that Higgs bosons 
decay only into SM particles.

In the sparticle sector
many different signatures have been studied~\cite{pauss} in the framework of 
the MSSM and mSUGRA. These studies include inclusive and exclusive 
signatures. Particular emphasis was given to 
the E$^{miss}_{T}$ and b--jet signatures.
In the following we briefly summarise the 
Higgs sector and  discuss some selected topics in sparticle searches.
%
%
\subsubsection {The MSSM Higgs sector}

The MSSM Higgs sector requires the existence of two Higgs doublets, 
resulting in five physical Higgs bosons~\cite{lowmassh}. Within this model,
at least one Higgs boson, the $h$, should have a mass smaller than 
m$_{h} \leq$ 125 GeV~\cite{newlowmassh}. The upper mass 
limit depends via radiative correction on the top quark mass,
the a priori unknown value of tan$\beta$ and also via a mixing parameter 
on the mass of the stop quark. One expects that  
such a MSSM Higgs boson should be found soon at LEP200 if 
nature has chosen a $\tan \beta$ value of smaller than 4.  
The masses of the other four Higgs bosons ${A}$, ${H^0}$ and ${H^\pm}$
are less constrained but should essentially be degenerate 
once their mass is larger than $\approx$ 200 GeV.

Current LHC studies show that the sensitivity to the MSSM Higgs sector is
somehow restricted. This is illustrated in figure 28, where
the  sensitivity of different signatures is  shown in a rather 
complicated two--dimensional  
multi--line contour plot.
\begin{figure}[htb]
\begin{center}
\includegraphics*[scale=0.6,bb=0 180 600 590 ]
{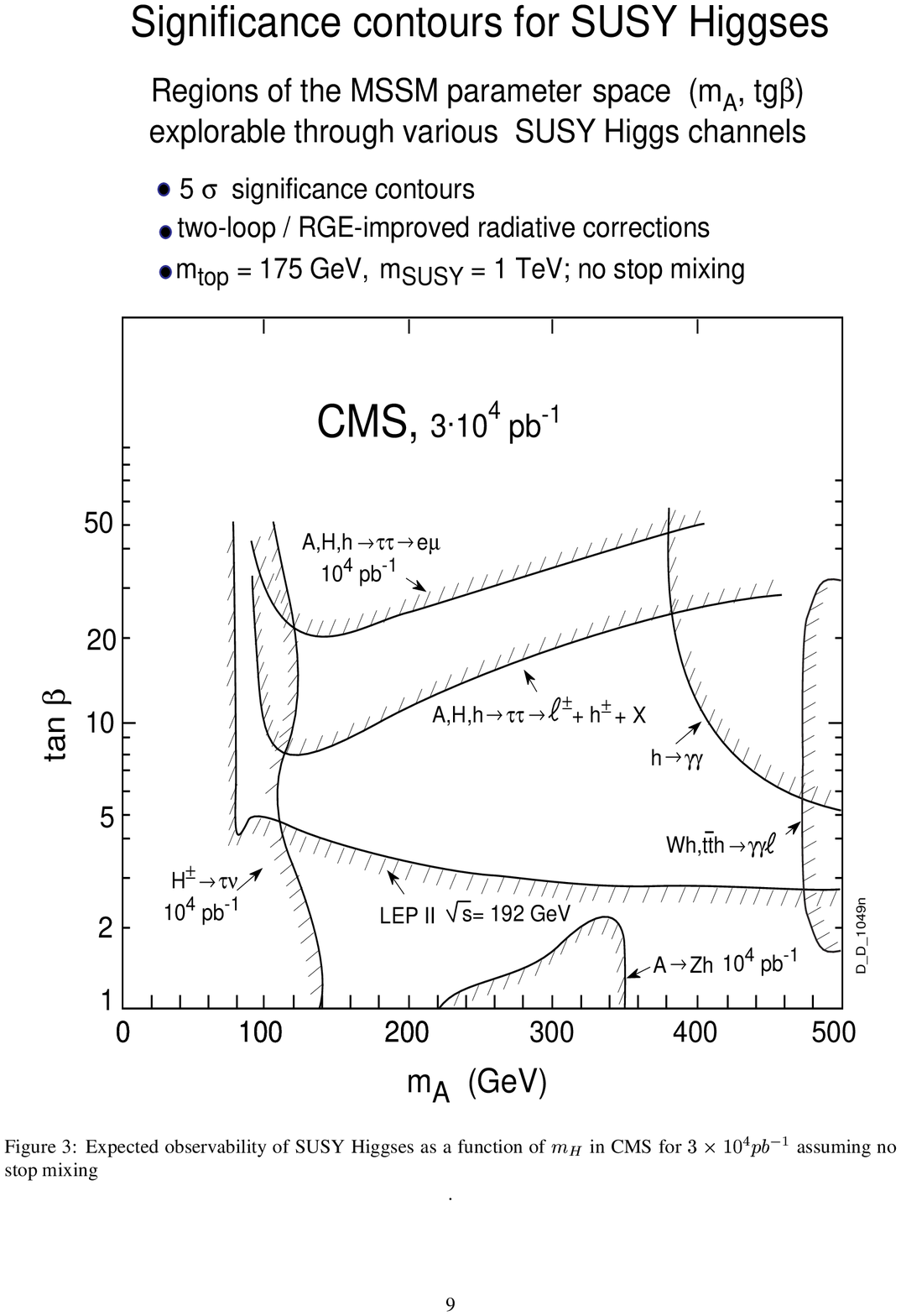}
\caption[fig28]
{CMS 5 sigma significance contour plot for the  MSSM Higgs 
sector in the m$_{A}$ -- $\tan\beta$ plane~\cite{cmshiggs1}.
Each curve indicates the sensitivity for a specific Higgs search mode.
No mixing in the stop sector is assumed.} 
\end{center}
\end{figure}
\paragraph{The lightest neutral Higgs $h$.}

For the lightest Higgs  the only established
signature appears to be the decay $h \rightarrow \gamma \gamma$. 
For large masses of m$_{A}$ (m$_{A} \geq$  400 GeV) one finds 
essentially
SM rates and  sensitivity. For much smaller masses of m$_{A}$, 
the branching ratio $h \rightarrow \gamma \gamma$ is
too small to observe a statistically significant signal.
The combination of the $h \rightarrow \gamma \gamma$ search 
with other $h$  decay modes, like $h \rightarrow  b\bar b$, might help
to enlarge the 5 sigma domain. A particular interesting aspect of an
inclusive higgs search in SUSY events with the decay 
$h \rightarrow  b\bar b$ decay will be discussed in the section
on sparticle searches. 
\paragraph{The heavy neutral Higgs Bosons $H^{0}, A^{0}$ and 
small values of $\tan \beta$.} 

Assuming $\tan \beta$ values below 4,  for some $H^{0}$ masses and decays, 
significant signals might be visible. 
For example, a $H^{0}$ with a mass close to 170 GeV  
appears to be detectable in the channel 
$H^{0} \rightarrow WW^{*} \rightarrow \ell \nu \ell \nu$. 
Other studies indicate the possibility to observe the decays
$H^{0} \rightarrow hh \rightarrow \gamma \gamma b \bar{b}$ and 
$A \rightarrow Z h \rightarrow \ell^{+} \ell^{-} b \bar{b}$ for 
masses between 200 -- 350 GeV.  
We will not go into further details here, as the relevance of such 
studies for low values of $\tan \beta$ depend very 
strongly on the forthcoming LEP200 results.
\paragraph{The heavy neutral Higgs bosons $H^{0}, A^{0}$ and
large values of $\tan \beta$.} 

For large values of $\tan \beta$, the Higgs production 
cross sections, especially the ones for $b\bar{b}H^{0}$ and $b\bar{b}A^{0}$, 
are much larger than for  the SM Higgs of similar mass.  
The only relevant Higgs decays are 
$H^{0}, A^{0} \rightarrow \tau \tau$ and 
$H^{0}, A^{0} \rightarrow b \bar{b}$.

Assuming large $\tan \beta$ values, the rare 
decay $A, H \rightarrow \mu \mu$ 
might show up as a resonance peak
above a large background\footnote{The branching ratio
is expected to be about a factor of 300 smaller than 
the one for the decay to $\tau\tau$, as it 
scales with $(m_{\mu}/m_{\tau})^{2}$.}.
Assuming excellent 
mass resolution in the $\mu \mu$ channel of about 
0.01--0.02$\times$m$_{Higgs}$ [in GeV],  a Higgs signal in this channel 
is detectable
for an integrated luminosity of 30 fb$^{-1}$
and $\tan \beta \geq 20$.
\paragraph{The charged Higgs $H^{\pm}$.} 

Depending only slightly on $\tan \beta$, the relevant 
charged Higgs decay modes are $H^{+} \rightarrow \tau^{+} \nu$
for masses below the $t \bar{b}$ threshold, and 
$H^{+} \rightarrow t \bar{b}$ above.  
Inclusive $t\bar{t}$ events might thus provide 
a good experimental signature for $H^{\pm}$ with a mass below 
$m_{top} - 10$ GeV. One has to search for $t\bar{t}$
events with isolated $\tau$ candidates which originate from the decay chain
$t\bar{t}  \rightarrow b W^{\pm} b H^{\pm}$ and 
$H^{\pm} \rightarrow \tau \nu$.

Another interesting process might be the 
production of a heavy $H^{\pm}$ in association with a top quark, 
$g b \rightarrow t H \rightarrow ttb \rightarrow WW bbb $. 
A parton level 
analysis of this channel~\cite{hplus} indicates the possibility 
to obtain $H^{\pm}$ mass peaks 
with reasonable signal--to--background ratios. 
\paragraph{Summary for the MSSM Higgs sector.} 

Figure 29 illustrates
the sensitivity of the ATLAS experiment
to various Higgs decay channels as discussed above~\cite{atlasmssmh}.
CMS reports very similar results.
These two--dimensional multi--line 
5 sigma (statistical) significance plots,
especially in the logarithmic version, indicate
sensitivity over almost the entire MSSM parameter space. 
\begin{figure}[htb]
\begin{center}
\epsfig{file=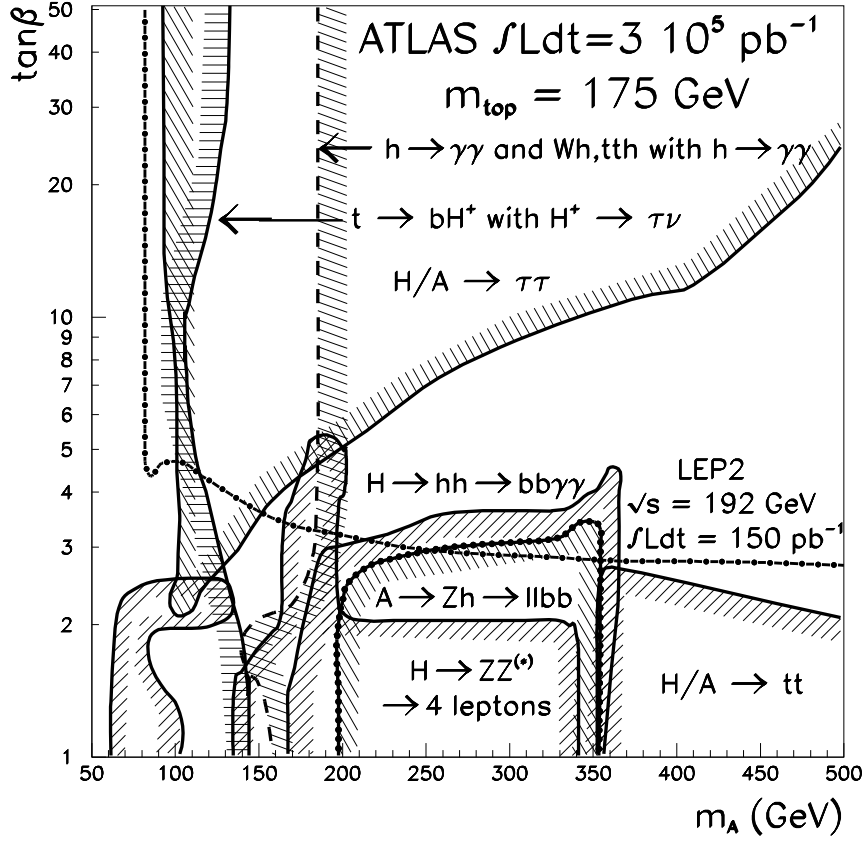,width=12.5cm}
\end{center}
\caption[fig29]
{Estimated ATLAS sensitivity (5 sigma) for 
MSSM Higgs searches, assuming 300 fb$^{-1}$~\cite{atlasmssmh}. 
The sensitivity of the different search signatures are shown 
in the m$_{A} - \tan \beta $ plane.}
\end{figure}
%
%
However, 
it is worth to remind the reader that 
only statistical errors and only decays into 
SM particles are included in the evaluation of the discovery potential.
As a consequence, the obtained 
curves, especially when extrapolated to larger integrated luminosities
and combined for ATLAS and CMS are doubtful. This is especially the 
case for channels like 
$WH \rightarrow \ell \nu b \bar{b}$ and 
for $H^{0},A^{0} \rightarrow \tau\tau$ where the proposed signatures 
suffer certainly from the very poor signal--to--background ratio.

%
%
\subsubsection {Searches for SUSY Particles}

The MSSM contains 124  free parameters including those of the SM.
That many free parameters do not offer a good
guidance for experimentalists. Additional  
assumptions to constrain the parameter space are therefore desirable.
The simplest approach is the so called mSUGRA 
(minimal Super--Gravity) model with 
only five additional new parameters ($m_{0}, m_{1/2}, \tan \beta, A^{0}$ 
and sign($\mu$) ).
This SUSY model is used for most of the simulation studies for which 
very advanced Monte Carlo programmes~\cite{isajet},~\cite{spythia},
and~\cite{pythia} exist.  
This pragmatic choice of one approach to investigate 
the discovery potential appears to be sufficient, 
as essentially all required detector features can be tested.

\subsubsection{mSUGRA predictions}

Signatures related to the MSSM and 
in particular to mSUGRA searches are
based on the consequences of R--parity conservation. R--parity is a 
multiplicative quantum number like ordinary parity. R--parity of 
the known SM
particles is +1, while it is --1 for sparticles. 
As a consequence, sparticles have to be produced in pairs.
Sparticles decay either directly or via 
some cascade processes to SM particles and the lightest supersymmetric
particle (LSP), which  is a neutral, massive and stable object with 
neutrino--like properties.
This LSP  should have been abundantly produced 
after the Big Bang and is an excellent candidate for cold 
dark matter. 
Usually one assumes the LSP to be the lightest neutralino 
$\tilde{\chi}^{0}_{1}$ which escapes detection.  
Large missing transverse energy is  thus the prime SUSY  
signature in collider experiments.

Within the mSUGRA model, the masses of 
sparticles are strongly related to the  universal fermion 
and scalar masses $m_{1/2}$ and $m_{0}$. The masses of the spin--1/2 
sparticles are directly related to $m_{1/2}$. 
One expects approximately the following mass hierarchy:
$\tilde{\chi}^{0}_{1} \approx 0.5\cdot m_{1/2} $,
$\tilde{\chi}^{0}_{2}\approx \tilde{\chi}^{\pm}_{1} \approx m_{1/2} $ and
$\tilde{g} \approx 3\cdot m_{1/2} $.
The masses of the spin--0 
sparticles are related to $m_{0}$ and $m_{1/2}$ and 
allow for some mass splitting between the ``left'' and ``right'' handed
scalar partners of the degenerated left and right handed fermions. 
One finds the following simplified mass relations: \\
$m(\tilde{q})$($\tilde{u}$, $\tilde{d}$, $\tilde{s}$, $\tilde{c}$ 
and $\tilde{b}$) 
$ \approx \sqrt{m_{0}^{2} + 6\cdot m_{1/2}^{2}}$,
$m(\tilde{\nu})_{left} \approx m(\tilde{\ell^{\pm}})_{left}
\approx \sqrt{m_{0}^{2} + 0.52\cdot m_{1/2}^{2}}$ and
$m(\tilde{\ell^{\pm}})_{right} \approx \sqrt{m_{0}^{2} +
0.15\cdot m_{1/2}^{2}}$.
The masses of the left and right handed stop quarks ($\tilde{t}_{\ell, r}$) 
can have a large splitting.
As a result, the right handed stop quark might be 
the lightest of all squarks.   
 
Following the above mass relations and using the known SUSY couplings, 
possible SUSY decays and the related signatures can be calculated.

As an  example we consider the $\tilde{\chi}^{0}_{2}$  decay 
$\tilde{\chi}^{0}_{2}\rightarrow \tilde{\chi}^{0}_{1}+ X$
with $X$ being:
$ \gamma^{*} Z^{*} \rightarrow \ell^{+}\ell^{-}$ or
$ h^{0} \rightarrow b \bar{b}$ or
$ Z  \rightarrow f \bar{f}$.
Other possible $\tilde{\chi}^{0}_{2}$ decay chains are 
$\tilde{\chi}^{0}_{2}\rightarrow \tilde{\chi}^{\pm(*)}_{1}+ \ell^{\mp} \nu$ 
and 
$\tilde{\chi}^{\pm(*)}_{1} \rightarrow \tilde{\chi}^{0}_{1} \ell^{\pm} \nu$
or 
$\tilde{\chi}^{0}_{2}\rightarrow \tilde{\ell}^{\pm} \ell^{\mp}$.
This example clearly demonstrates the complexity of sparticle signatures.

For higher sparticle masses, even more decay channels
might open up. It is thus not possible to define all search strategies
a priori. Furthermore, possible 
unconstrained mixing angles between neutralinos lead to a
model dependent search strategy for squarks and gluinos, as 
will be discussed below.

The discovery potential is
usually given in the $m_{0}$--$m_{1/2}$ parameter space. 
Despite the model dependence, it allows to 
compare the sensitivity  of  different signatures. 
Having various proposed methods, the resulting 
figures are -- as in the Higgs sector -- rather complicated and require 
some time for appreciation.   
A typical example is shown in figure 30~\cite{baer96}, 
where the different 
curves indicate the LHC sensitivity for different signatures 
and different sparticles. It is usually assumed that the maximum 
information about SUSY can be extracted in regions which are covered 
by many different signatures. 
The meaning of the various curves should become clear 
in the following section.  
\begin{figure}[htb]
\begin{center}
\includegraphics*[scale=0.7,bb=40 200 600 670 ]
{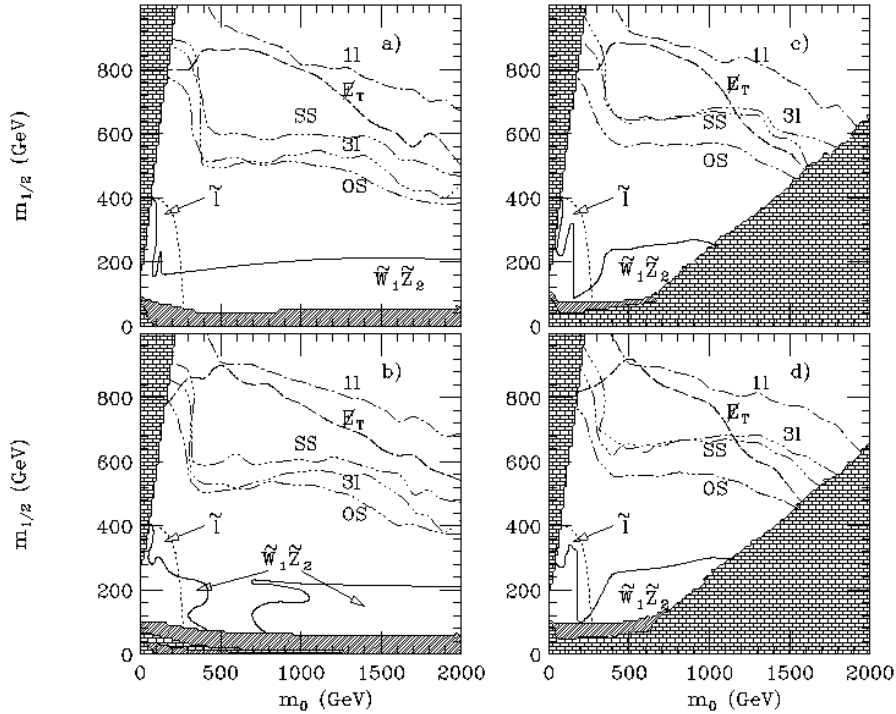}
\caption[fig30]
{Expected mSUGRA sensitivity of various signatures 
in the $m_{0} - m_{1/2}$ plane at the LHC,
assuming an integrated luminosity of 
10 fb$^{-1}$~\cite{baer96}. The different curves indicate 
the expected sensitivity for SUSY events with n leptons ($\ell$) 
and for events with lepton pairs with same charge (SS) and 
opposite charge (OS).}   
\end{center}
\end{figure}
%
%
%

\subsubsection{Squark and Gluino searches}

The cross--section for strongly interacting sparticles  are  large at
the LHC, as can be seen in figure 31~\cite{baer96}.
For example the pair production cross--section of 
squarks and gluinos with a mass of $\approx$ 1 TeV has been estimated 
to be as large as 1 pb resulting in 
10$^{4}$ produced SUSY events for 10 fb$^{-1}$. 
This high rate,
combined with the possibility to observe many different 
decay modes, could turn LHC into a ``SUSY--factory''.
\begin{figure}[htb]
\begin{center}
\includegraphics*[scale=0.6,bb=-38 321 360 628]
{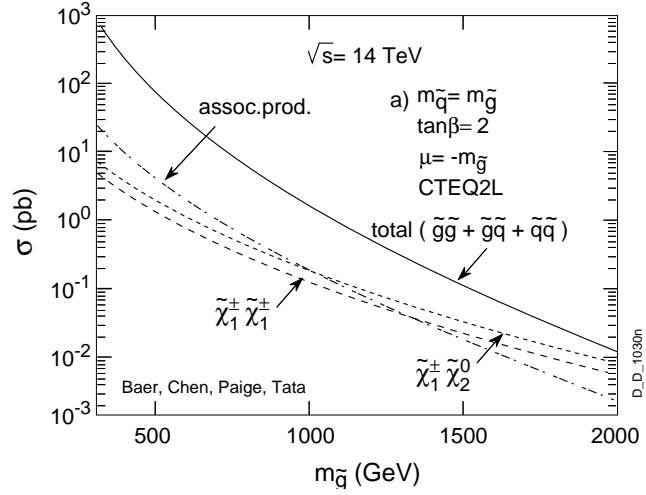}
\caption[fig31]
{Cross-sections for sparticle production at the LHC~\cite{baer96}.} 
\end{center}
\end{figure}

Depending on the SUSY model parameters, a large variety of 
massive squark and gluino decay channels and signatures might exist.
A search for squarks and gluons 
at the LHC should consider the various 
signatures resulting from the following decay channels:
\begin{itemize}
\item $\tilde{g} \rightarrow \tilde{q}\bar{q}$ including 
$\tilde{g} \rightarrow \tilde{t} \bar{t}$ 
\item $\tilde{q} \rightarrow \tilde{\chi}^{0}_{1} q$ ~~~or~~~
$\tilde{q} \rightarrow \tilde{\chi}^{0}_{2} q$  ~~~or~~~
$\tilde{q} \rightarrow \tilde{\chi}^{\pm}_{1} q'$
\item $\tilde{\chi}^{0}_{2} \rightarrow \ell^{+}\ell^{-}\tilde{\chi}^{0}_{1} $ 
 ~~~or~~~ $\tilde{\chi}^{0}_{2} \rightarrow Z^{0} \tilde{\chi}^{0}_{1}$  ~~~or~~~
$\tilde{\chi}^{0}_{2} \rightarrow h^{0} \tilde{\chi}^{0}_{1}$
\item $\tilde{\chi}^{\pm}_{1} \rightarrow 
\tilde{\chi}^{0}_{1} \ell^{\pm}\nu$   ~~~or~~~
$\tilde{\chi}^{\pm}_{1} \rightarrow W \tilde{\chi}^{0}_{1}$.
\end{itemize}  
The various decay channels 
can be separated into at least three distinct event signatures.
\begin{itemize}
\item Multi--jets + missing transverse energy; these events 
should be spherical in the plane transverse to the beam.
\item Multi--jets + missing transverse energy + n(=1,2,3,4) 
isolated high p$_{T}$ leptons; these leptons originate 
from cascade decays of charginos and neutralinos.
\item
Multi--jets + missing transverse energy + lepton pairs of the same charge; 
such events can originate from 
$\tilde{g}\tilde{g} \rightarrow \tilde{u} \bar{u} \tilde{d} \bar{d}$
with subsequent decays of the squarks to 
$\tilde{u} \rightarrow \tilde{\chi}^{+} d$ 
and $\tilde{d} \rightarrow \tilde{\chi}^{+} u$ 
followed by leptonic chargino decays  
$\tilde{\chi}^{+} \rightarrow \tilde{\chi}^{0}_{1} \ell^{+} \nu$.  
\end{itemize}
The observation and detailed analysis
of  different types of SUSY events might allow
the discovery of many sparticles and should 
help to measure some of the SUSY parameters.

A search strategy for squarks and gluinos 
at the LHC would select 
jet events with large visible transverse mass and missing 
transverse energy. Such events can then be classified according to
the number of isolated high p$_{T}$ leptons including the lepton
flavour and charge relation. Once an excess above 
SM backgrounds is observed, one would try to interpret the 
 events and measured cross--section(s) in terms of
$\tilde{g}, \tilde{q}$ masses and decay modes for various SUSY models. 
Concerning the SM background processes,
the largest backgrounds originate mainly from 
$W+$jet(s), $Z+$jet(s) and $t\bar{t}$ events. 
Using this approach, very encouraging
signal--to--background ratios, combined with sizable signal cross--sections, 
are obtainable for a large range of squark and gluino masses.
\begin{figure}[htb]
\begin{center}
\includegraphics*[scale=0.5,bb=40 50 600 580 ]
{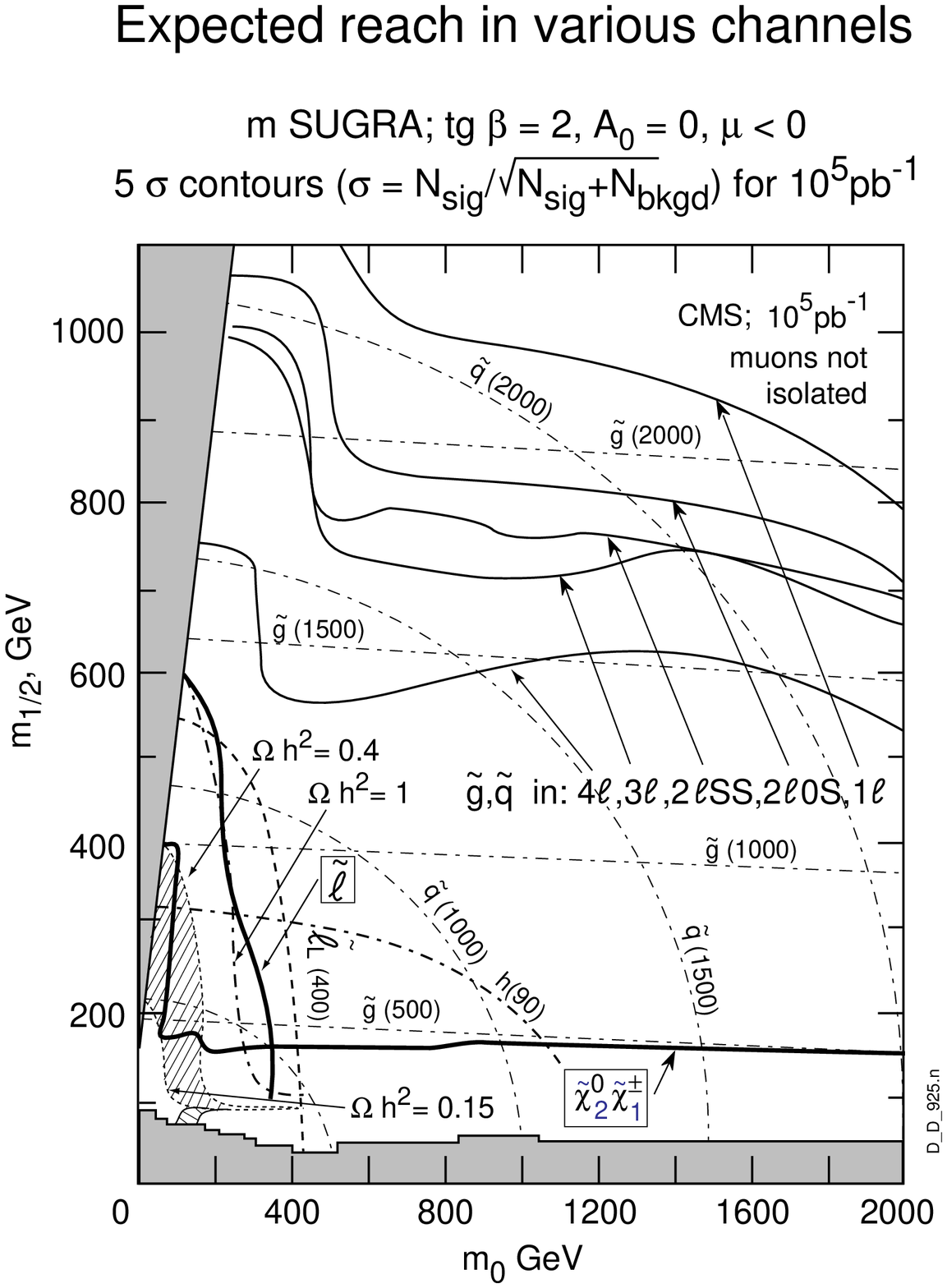}
\caption[fig32]
{Expected  CMS sensitivity 
for squarks and gluinos, sleptons and for 
$\tilde{\chi}^{0}_{2}\tilde{\chi}^{\pm}_{1}$
in the $m_{0}-m_{1/2}$ plane, assuming 100 fb$^{-1}$~\cite{cms98006}.
The different solid lines show the 
expected 5 sigma signal, estimated from S/$\sqrt{S+B_{SM}}$, 
coverage domain for the various signatures
using isolated high p$_{T}$ leptons. The dashed--dotted lines indicate the 
corresponding squark and gluino masses.}
\end{center}
\end{figure}
The simulation results of  
such studies indicate, as shown in figure 32,
that LHC experiments are sensitive to 
squark and gluinos masses up to about 2 TeV, 
assuming 100 fb$^{-1}$~\cite{cms98006}. 
Figure 32 also illustrates that detailed studies of branching ratios 
using the different decay chains are 
possible up to squark or gluino masses of about 1.5 TeV, where 
significant signals can be observed for many different channels.   
Another consequence of the expected large signal cross--sections
is the possibility that at the  LHC start--up, for an integrated 
luminosity of only $\approx$ 100 pb$^{-1}$, 
squarks and gluinos up to masses of about 600 to 700 GeV can be discovered,
which is well beyond the most optimistic 
Tevatron Run III accessible mass range.

Given this exciting squark and gluino discovery potential for 
many different channels, one has to keep in mind that
all potential signals depend strongly on the  understanding  
of the various background processes and thus on the detector systematics.
A thorough study of backgrounds including the shapes of
background distributions is therefore mandatory. 
In particular, the requirements of very efficient lepton identification 
and good missing transverse energy measurement demand for a
well understood detector response.
This will require certainly some time, 
given the complexity of the LHC detectors.

Our discussion of the  SUSY discovery potential 
has illustrated the sensitivity of the proposed ATLAS and CMS experiments.
The next step after a discovery is the determination  
of SUSY parameters, thus obtaining a deeper insight of the 
underlying theory.   

\begin{figure}[htb]
\begin{center}
\includegraphics*[scale=0.6,bb=-26 -130 373 259 height=1. cm,width=8.cm]
{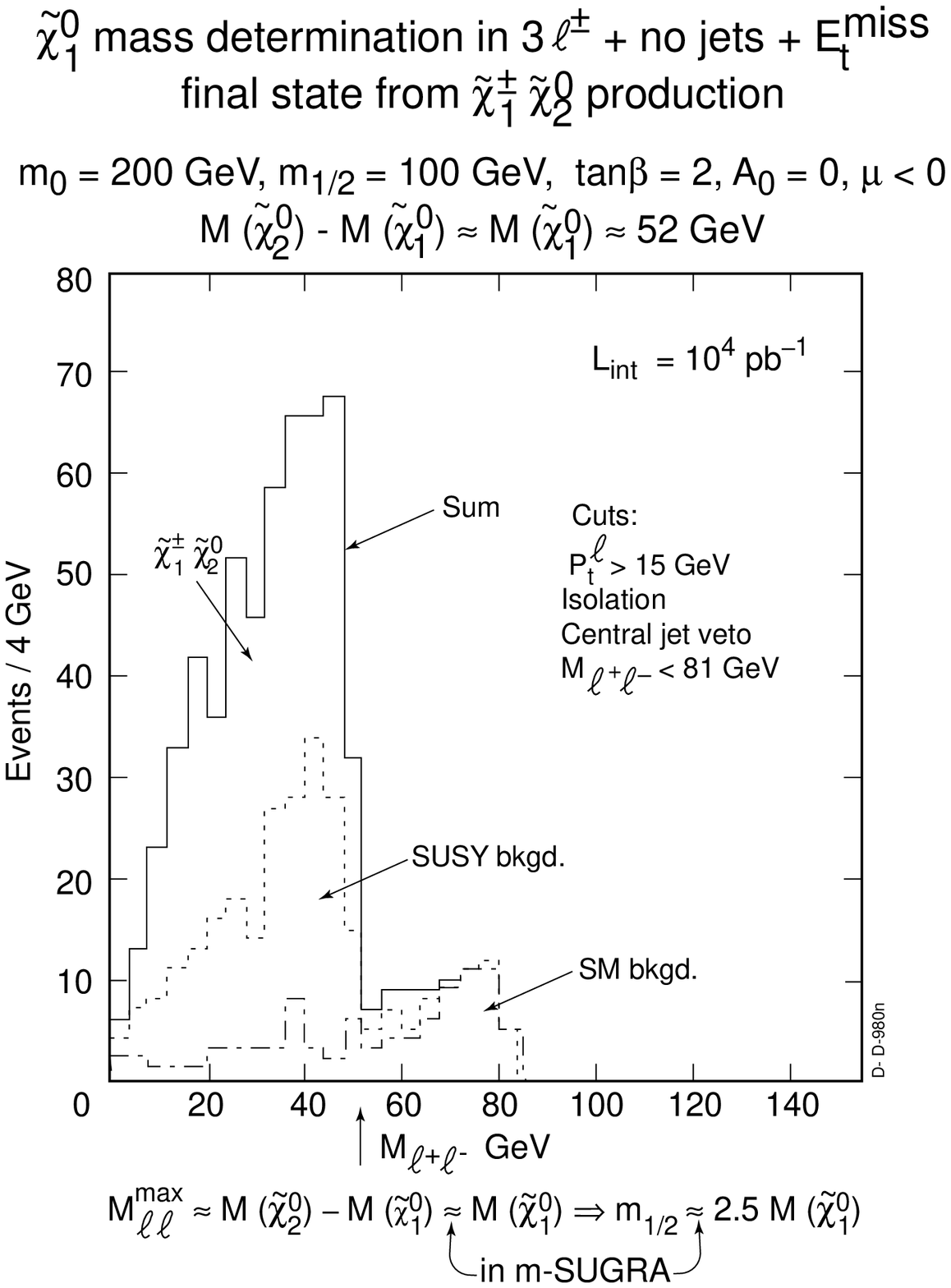}
\caption[fig33]
{Expected dilepton mass distribution in CMS for L=10 fb$^{-1}$ and 
for trilepton events from 
$\tilde{\chi}^{0}_{2}\tilde{\chi}^{\pm}_{1}$ decays~\cite{cms98006}.
The upper edge in the distribution at about 
50 GeV corresponds to the kinematic 
limit in the decay 
$\tilde{\chi}^{0}_{2} \rightarrow \ell^{+}\ell^{-}\tilde{\chi}^{0}_{1}$
and is thus sensitive to the mass difference 
of $\tilde{\chi}^{0}_{2} - \tilde{\chi}^{0}_{1}$.}
\end{center}
\end{figure}
 
The
production and decays of $\tilde{\chi}^{0}_{2}\tilde{\chi}^{\pm}_{1}$
provide enough rates for masses below 200 GeV and should allow, 
as shown in figure 33, to measure accurately the dilepton mass 
distribution and their relative p$_{T}$ spectra.
The mass distribution and especially the edge 
in the mass distribution has been shown to be sensitive
to the mass difference between the two neutralinos involved in the decay.
 
In contrast to the rate limitations of weakly produced  
sparticles at the LHC, detailed 
studies of clean squark and gluino events are expected 
to reveal more information.  
One finds that the large rate for many 
distinct event signatures allows to measure masses and mass 
ratios for several SUSY particles, 
produced in cascade decays of squarks and gluons. 
Many of these ideas have been discussed at the 1996 
CERN SUSY--Workshop~\cite{cernth96}.
An especially interesting proposal is
the detection of the Higgs boson, $h$, via the decay chain 
$\tilde{\chi}^{0}_{2} \rightarrow \tilde{\chi}^{0}_{1} h \rightarrow 
\tilde{\chi}^{0}_{1} b \bar{b}$. 
The simulated mass distribution for $b \bar{b}$--jets 
in events with large missing transverse energy is shown 
in figure 34~\cite{cms98006}. 
Higgs mass peaks above background are found 
for various choices of $\tan \beta$ and $m_{0}, m_{1/2}$.  

\begin{figure}[htb]
\begin{center}
\includegraphics*[scale=1.2,bb=0 300 550 750 height=10. cm,width=14.cm]
{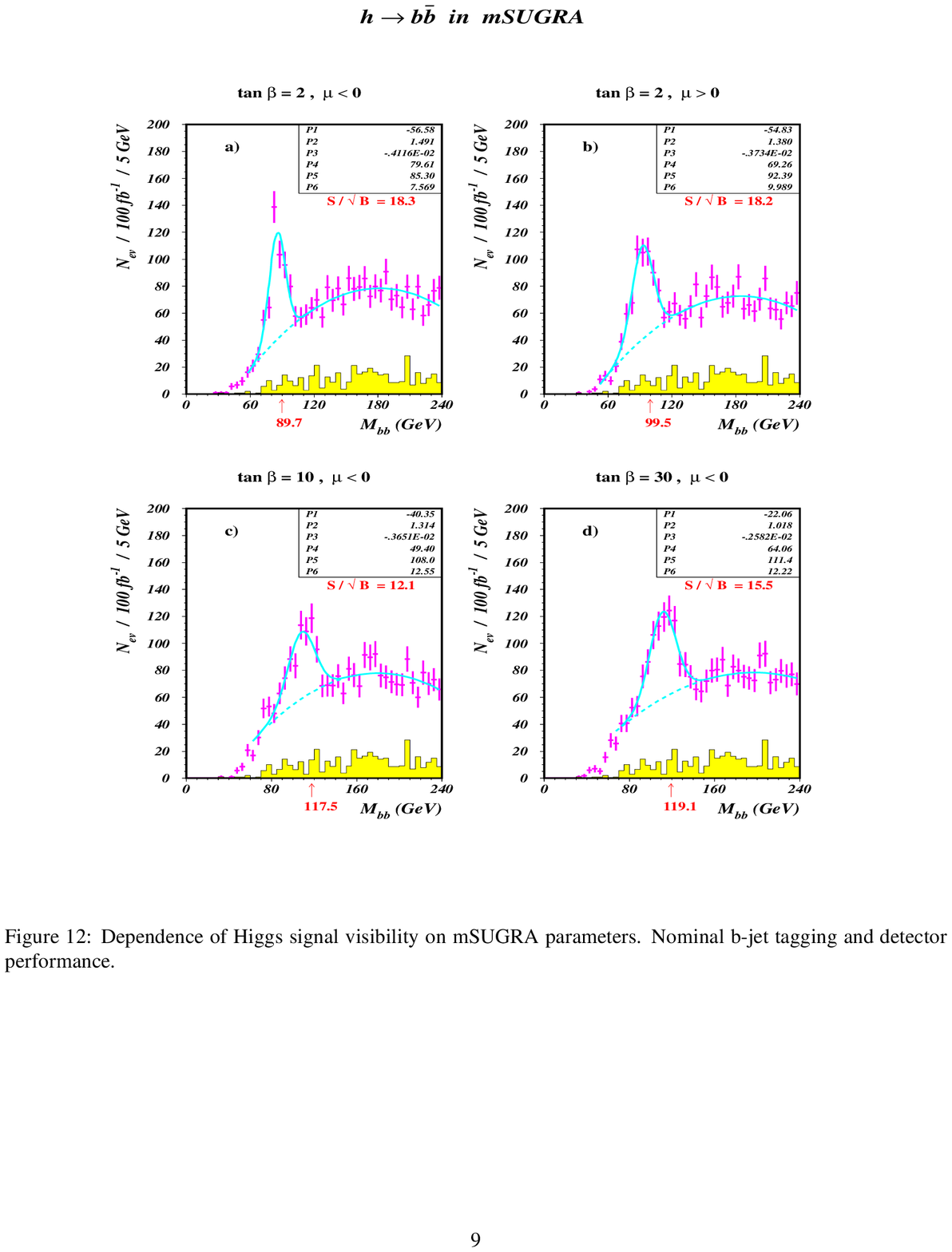}
\caption[fig34]
{Higgs signals in squark and gluino events,
reconstructed from the invariant mass of $h \rightarrow b \bar{b}$ 
and assuming 100 fb$^{-1}$ in the CMS experiment~\cite{cms98006}.}
\end{center}
\end{figure}
 
Another interesting approach to determine the 
SUSY mass scale has been suggested in~\cite{atlas107}.
The idea is to define an effective mass,  
using the scalar p$_{T}$ sum of the four 
jets with the largest transverse energy,
plus the missing transverse energy of the event. 
This effective mass shows an approximately linear 
relation with the underlying SUSY mass--scale, defined 
as the minimum of the squark or gluino mass.

The possibility to extract model parameters within the 
mSUGRA framework was illustrated using the following method:
for each point in the parameter space a set of experimental 
constraints on sparticle masses is derived and a fit 
to parameters of mSUGRA is performed. 
For example, assuming m$_{0}$ = m$_{1/2}$ = 400 GeV, 
A$_{0}$ = 0 , $\tan \beta$ = 2 
and sign($\mu$) = +, one obtains for 30 fb$^{-1}$ the following 
sensitivity: m$_{0}$ = 400$\pm$100 GeV, m$_{1/2}$ = 400$\pm$10 GeV 
and $\tan \beta$ = 2$\pm$0.8.

\section{Concluding Remarks}

The LHC is currently the only realistic possibility to reach the 
TeV energy range in constituent--constituent scattering; 
this energy range is expected to be rich in discoveries.
 
A large international community is working on the realization of the LHC
experimental programme. The two large general--purpose detectors 
ATLAS and CMS have moved from the R\&D to preproduction and in 
certain areas to the production phase. 
The concept of both detectors was driven by physics considerations 
using SM and beyond the SM processes, which were used to optimise the 
final detector design.  

It was already pointed out that,
although the most exciting discoveries will be those of totally 
unexpected new particles or phenomena, one can only
demonstrate the discovery potential of the proposed experiments 
using predicted new particles. However, the experiments 
designed under these considerations should also allow the 
discovery of whatever new phenomena might occur in 
multi--TeV pp collisions.

\vskip 1.0cm

{\bf \large Acknowledgements}

{\small One of us (F.P.) would like to thank Tom Ferbel 
for the invitation to this school, which was in all respects very 
interesting and superbly organised.
}
 
\end{document}